\documentclass[twocolumn,pre,eqsecnum,showpacs]{revtex4}

\usepackage{graphicx}
\usepackage{epsfig}
\usepackage{dcolumn}
\usepackage{amsfonts}
\usepackage{amsmath}
\usepackage{amssymb}
\usepackage{bm}
\usepackage{longtable}
\usepackage{latexsym}

\begin{document}

%shortcuts
\newcommand{\tens}[1]{\tensor{#1}}
\newcommand{\brm}[1]{\bm{{\rm #1}}}

\title{Multifractal properties of resistor diode percolation}

\author{Olaf Stenull\footnote{Present address: Department of Physics and Astronomy, University of Pennsylvania, Philadelphia, PA 19104-6396, USA}}
\author{Hans-Karl Janssen}
\affiliation{
Institut f\"{u}r Theoretische Physik 
III\\Heinrich-Heine-Universit\"{a}t\\Universit\"{a}tsstra{\ss}e 1\\
40225 D\"{u}sseldorf\\
Germany
}

\date{\today}

\begin{abstract}
Focusing on multifractal properties we investigate electric transport on random resistor diode networks at the phase transition between the non-percolating and the directed percolating phase. Building on first principles such as symmetries and relevance we derive a field theoretic Hamiltonian. Based on this Hamiltonian we determine the multifractal moments of the current distribution that are governed by a family of critical exponents $\{ \psi_l \}$. We calculate the family $\{ \psi_l \}$ to two-loop order in a diagrammatic perturbation calculation augmented by renormalization group methods.
\end{abstract}

\maketitle

\section{Introduction}
Percolation~\cite{bunde_havlin_91_etc} has been of tremendous interest to diverse scientific disciplines for more than two decades. On one hand, the percolation problem is almost trivially simple to state, on the other hand, it has an abundance of fascinating features. Although percolation is purely geometric in nature, it reflects many of the concepts of critical phenomena. The percolation transition is the prototype of a geometrical phase transition. In this respect, the role of percolation may be compared to that of the Ising model in conventional critical phenomena. To date, percolation represents one of the best studied areas of statistical physics and yet it is a highly vivid and fascinating area of modern research. 

Directed percolation (DP)~\cite{hinrichsen_2000} is an anisotropic variant of ordinary isotropic percolation (IP) in which an effect or activity can percolate only in a given preferred direction. If the preferred (longitudinal) direction is viewed as time, DP can be interpreted as a dynamic process. In the dynamic interpretation, DP represents one of the most prominent universality classes of nonequilibrium phase transitions: the DP universality class is the generic universality class for phase transitions between an active and an absorbing inactive state~\cite{janssen_81,grassberger_82}.

Perhaps, DP is the simplest model resulting in branching self-affine objects. It has many potential applications, including fluid flow through porous media under gravity, hopping conductivity in a strong electric field~\cite{vanLien_shklovskii_81}, crack propagation~\cite{kertez_vicsek_80}, and the propagation of surfaces at depinning transitions in one dimension~\cite{depinning}. Moreover, it is related to epidemics with recovery~\cite{grassberger_85} and self-organized critical models~\cite{soc}.

In this paper we discuss multifractality~\cite{reviews_multifractality} in DP. The notion of multifractality describes the situation that an exhaustive characterization of the distribution of a local physical quantity requires the introduction of an infinite set of independent critical exponents. This means that the distribution is not controlled by a single or several length scales, but rather by an infinite hierarchy of such. Systems in which multifractality has been observed include turbulence~\cite{mandelbrot_74}, diffusion near fractals~\cite{cates_witten_87} and polymers~\cite{vonFerber_holovatch}, electrons in disordered media~\cite{wegner_87}, polymers in disordered media~\cite{gersappe_etal_91}, random ferromagnets~\cite{ludwig_87}, chaotic dissipative systems~\cite{halsey_etal_86}, and heartbeat~\cite{ivanov_99_etc}.

Since the 1980s several publications have appeared that have addressed multifractality in IP by studying random resistor networks (RRNs)~\cite{rammal_etal_pra_85,rammal_etal_prl_85,arcangelis_etal_85,park_harris_lubensky_87,stenull_janssen_epl_2000,stenull_2000,stenull_janssen_2001}. A RRN is a simple bond percolation model on a $d$-dimensional hypercubic lattice where bonds between nearest neighboring sites are occupied with probability $p$ with a resistor or empty with probability $1-p$. In a typical setup one has two leads positioned at two connected lattice sites $x$ and $x^\prime$ at which a fixed external current $I$ is injected and, respectively, extracted. It is well established that the distribution of bond currents in RRN is multifractal. This multifractality is accessible by experiments or simulations via the moments of the distribution that are [up to technical details, cf. Eq.~(\ref{multiMoment})] given by 
\begin{eqnarray}
M_I^{(l)} \left( x, x^\prime \right) \sim \sum_{\underline{b}} \left( I_{\underline{b}}/I \right)^{2l} \, .
\end{eqnarray}
Here $I_{\underline{b}}$ denotes the current flowing through bond $\underline{b}$ and the sum is taken over all bonds. Some of these multifractal moments are corresponding to quantities that have a particularly prominent role in percolation theory. For example, $M_I^{(0)}$, $M_I^{(1)}$, $M_I^{(2)}$, and $M_I^{(\infty )}$ are proportional to the number of bonds on the backbone (bonds which carry nonzero current), the total resistance, the noise (second cumulant of the resistance fluctuations), and the number of the red bonds (bonds which carry the full current). At the percolation transition, the moments are described by a power law in the separation between the leads, 
\begin{eqnarray}
M_I^{(l)} \left( x, x^\prime \right) \sim \big| x - x^\prime \big|^{\psi^{\text{IP}}_l/\nu} \, ,
\end{eqnarray}
where $\nu$ is the correlation length exponent of the IP universality class. The $\psi^{\text{IP}}_l$ are referred to as multifractal exponents. The multifractality of the current distribution manifests itself in the fact that the $\psi^{\text{IP}}_l$, when understood as a function of the index $l$, are not related to each other in a linear or affine fashion.

To our knowledge the issue of multifractality in DP has not been addressed hitherto. Here, we present such a study. By employing our real world interpretation of Feynman diagrams~\cite{stenull_janssen_oerding_99,janssen_stenull_oerding_99,janssen_stenull_99,stenull_janssen_epl_2000,stenull_2000,stenull_janssen_2001,janssen_stenull_prerapid_2001,stenull_janssen_oerding_2001,stenull_janssen_pre_2001_nonlinear,stenull_janssen_jsp_2001,janssen_stenull_pre_2001_vulcanization,janssen_stenull_pre_2001_continuum} as well as our concept of master operators~\cite{stenull_janssen_epl_2000}, we find that the moments of the current distribution in DP are governed by a family $\{ \psi_l\}$ of multifractal exponents analogous to $\{ \psi^{\text{IP}}_l \}$. For example, if the multifractal moments are measured in the preferred time-like direction between two leads at $x = (x_\perp =0, t)$ and $x^\prime = (x_\perp^\prime =0, t^\prime)$, then we find that
\begin{eqnarray}
M_I^{(l)} \left( x, x^\prime \right) \sim \left( t -t^\prime \right)^{\psi_l/\nu_\parallel} \ ,
\end{eqnarray}
where $\nu_\parallel$ is the longitudinal correlation length exponent of the DP universality class. We calculate the family $\{ \psi_l\}$ to second order in an $\varepsilon$-expansion about the upper critical dimension 5.

Brief account of our results has been given in Ref.~\cite{stenull_janssen_epl_2001}. The aim of the present paper is to present our work in some detail. Its remainder is organized as follows: Section~\ref{model} is devoted to modeling issues. We start by briefly reviewing random resistor diode networks and Ohm's and Kirchhoff's laws. Then, static noise is introduced in the networks and corresponding noise cumulants are defined. These noise cumulants are important because they are closely related to the multifractal moments that are not accessible by our methods directly. Then, we set up a generating function for the noise cumulants by employing the replica trick. The replication leads to an effective Hamiltonian that we refine into a field theoretic Landau-Ginzburg-Wilson functional $\mathcal{H}$. We conclude Sec.~\ref{model} by analyzing the relevance of various coupling constants appearing in $\mathcal{H}$. We show that the noise cumulants are associated with dangerously irrelevant couplings. Section~\ref{RGA} contains the core of our renormalization group analysis. Its main content is the calculation of the generating function for the noise cumulants by employing renormalized field theory. It starts with determining the diagrammatic elements of our perturbation calculation upon recasting $\mathcal{H}$ into the form of a dynamic functional. Next, we demonstrate how we incorporate the dangerously irrelevant noise couplings into our diagrammatic calculation via operator insertions. We show that the operators associated with the noise cumulants are master operators. We derive the scaling behavior of the generating functions. Then, the scaling behavior of the noise cumulants and multifractal moments is readily extracted via taking derivatives. Finally, our result for $\{ \psi_l\}$ is stated and several consistency checks are provided. Concluding remarks are given in Sec.~\ref{conclusions}. Technical details are relegated to two Appendices.

\section{The Model}
\label{model}

\subsection{Random resistor diode networks}
Our approach is based on a model which captures both, IP and DP, {\em viz.} the random resistor diode network (RDN) introduced by Redner~\cite{red_81&82a,red_83,perc}. We start by reviewing substantial features of the RDN. A RDN consists of a $d$-dimensional hypercubic lattice in which nearest-neighbor sites are connected by a resistor, a positive diode (conducting only in a distinguished direction), a negative diode (conducting only opposite to the distinguished direction), or an insulator with respective probabilities $p$, $p_{+}$, $p_{-}$, and $q=1-p-p_{+}-p_{-}$. The three dimensional phase diagram (pictured as a tetrahedron spanned by the four probabilities) comprises a nonpercolating and three percolating phases. The percolating phases are isotropic, positively directed, or negatively directed. Between the phases there are surfaces of continuous transitions. All four phases meet along a multicritical line, where $0\leq r:=p_{+}=p_{-}\leq 1/2$ and $p=p_{c}(r)$. On the entire multicritical line, i.e., independently of $r$, one finds the scaling properties of usual isotropic percolation ($r=0$). For the crossover from IP to DP see, e.g., Ref.~\cite{janssen_stenull_2000}.
 
\subsection{Kirchhoff's laws}
To be specific we choose ${\rm{\bf n}} = 1/\sqrt{d} \left( 1, \dots , 1 \right)$ as the preferred direction. We assume that the bonds $\underline{b}_{\langle i,j \rangle}$ between two nearest neighboring lattice sites $i$ and $j$ are directed so that $\underline{b}_{\langle i,j \rangle} \cdot {\rm{\bf n}} > 0$. We suppose that the directed bonds obey the non-linear generalization of Ohm's law
\begin{eqnarray}
\label{avoid}
\sigma_{\underline{b}_{\langle i,j \rangle}} \left( V_{\underline{b}_{\langle i,j \rangle}} \right) V_{\underline{b}_{\langle i,j \rangle}} = I_{\underline{b}_{\langle i,j \rangle}}\, ,
\end{eqnarray}
where $V_{\underline{b}_{\langle i,j \rangle}} = V_j - V_i$ is the voltage drop over the bond between sites $j$ and $i$ and $I_{\underline{b}_{\langle i,j \rangle}}$ denotes the current flowing from $j$ to $i$. In the following we drop the subscript $\langle i,j \rangle$ whenever there is no risk of confusion. The bond conductances $\sigma_{\underline{b}} = \varsigma_{\underline{b}} \, \gamma_{\underline{b}}$ are equally and independently distributed random variables. The $\gamma_{\underline{b}}$ take on the values $1$, $\theta \left( V \right)$, $\theta \left( -V \right)$, and $0$ with respective probabilities $p$, $p_+$, $p_-$, and $q$. $\theta$ denotes the Heaviside function. The nature of the $\varsigma_{\underline{b}}$ will be specified below. Note that the diodes are idealized: under forward-bias voltage they behave as ``ohmic'' resistors whereas they are insulating under backward-bias voltage.  To prevent confusion, we point out that the round brackets in Eq.~(\ref{avoid}) contain the functional argument of the bond conductance. Because the bond conductance depends on the voltage drop over that bond only via a Heaviside function and $\mbox{sign} \left( V_{\underline{b}} \right) = \mbox{sign} \left( I_{\underline{b}} \right)$ we may write $\sigma_{\underline{b}} \left( V_{\underline{b}} \right) = \sigma_{\underline{b}} \left( I_{\underline{b}} \right)$.

Assume that an external current $I$ is injected at $x$ and extracted at $x^\prime$. It is understood that $x$ and $x^\prime$ are connected. The power dissipated on the network is by definition 
\begin{eqnarray}
P=I \left[ V_x - V_{x^\prime} \right] \, .
\end{eqnarray}
Using Ohm's law it may be expressed entirely in terms of the voltages as
\begin{eqnarray}
\label{power1}
P &=& R_+(x ,x^\prime)^{-1} \left[ V_x - V_{x^\prime} \right]^2 
\nonumber \\
&=& \sum_{\underline{b}} \sigma_{\underline{b}} \left( V_{\underline{b}} \right) V_{\underline{b}}^2 = P \left( \left\{ V \right\} \right) \, .
\end{eqnarray}
The sum is taken over all current carrying bonds (the backbone) between $x$ and $x^\prime$ and $\left\{ V \right\}$ denotes the corresponding set of voltages. $R_+(x ,x^\prime)$ stands for the macroscopic resistance when $I$ is inserted at $x$ and withdrawn at $x^\prime$. Similarly one defines $R_-(x ,x^\prime)$ as the macroscopic resistance when $I$ is inserted at $x^\prime$ and withdrawn at $x$. The two quantities, however, are related by $R_+(x ,x^\prime) = R_-(x^\prime ,x)$. From the power one obtains Kirchhoff's first law
\begin{eqnarray}
\label{kirchhoff}
\sum_{\langle j \rangle} \sigma_{\underline{b}_{\langle i,j \rangle}} \left( V_{\underline{b}_{\langle i,j \rangle}} \right) V_{\underline{b}_{\langle i,j \rangle}} = \sum_{\langle j \rangle} I_{\underline{b}_{\langle i,j \rangle}} = - I_i 
\end{eqnarray}
as a consequence of the variation principle
\begin{eqnarray}
\label{variationPrinciple1}
\frac{\partial}{\partial V_i} \left[ \frac{1}{2} P \left( \left\{ V \right\} \right) + \sum_k I_k V_k \right] = 0 \, .
\end{eqnarray}
The summation in Eq.~(\ref{kirchhoff}) extends over the nearest neighbors of $i$ and $I_i$ is given by $I_i = I \left( \delta_{i,x} - \delta_{i,x^\prime} \right)$.

Alternatively to Eq.~(\ref{power1}), the power can be expressed in terms of the currents as
\begin{eqnarray}
\label{power2}
P = R_+ \left( x ,x^\prime \right) I^2 = \sum_{\underline{b}} \rho_{\underline{b}} \left( I_{\underline{b}} \right) I_{\underline{b}}^2 = P \left( \left\{ I \right\} \right) \, ,
\end{eqnarray}
with $\left\{ I \right\}$ denoting the set of currents flowing through the individual bonds. $\rho_{\underline{b}} = \sigma_{\underline{b}}^{-1}$ denotes a bond resistance. It is understood that $\rho_{\underline{b}} \left( I_{\underline{b}} \right) I_{\underline{b}}^2 =0$ whenever $\sigma_{\underline{b}} \left( I_{\underline{b}} \right) = 0$. Kirchhoff's second law, saying that the voltage drops along closed loops vanish, can be stated in terms of the variation principle
\begin{eqnarray}
\label{variationPrinciple2}
\frac{\partial}{\partial I^{(l)}} P \left( \left\{ I^{(l)} \right\} , I \right) = 0 \, ,
\end{eqnarray}  
i.e., there are no independent loop currents $I^{(l)}$ circulating around a complete set of independent closed loops.

As customary in dealing with electric networks, Kirchhoff's equations may be exploited to calculate the total resistance $R_+$. Of course we are not primarily interested in any particular random configuration $C$ of the diluted network but rather in an average $\langle \cdots \rangle_C$ over all these configurations. This average, however, requires a little caution because the resistance between sites not connected by any conducting path is infinite. Therefore, we will exclusively consider those sites  $x$ and $x^\prime$ known to be connected by such a path. In practice this is done by utilizing an indicator function $\chi_+ (x ,x^\prime)$ that takes the value one if $x$ and $x^\prime$ are positively connected, i.e., if $I$ can percolate from $x$ to $x^\prime$, and zero otherwise. Note that $\langle \chi_+ (x ,x^\prime) \rangle_C = \langle \chi_- (x^\prime ,x) \rangle_C$ is nothing more than the usual DP correlation function. With help of the indicator function the average resistance, or more general the $l$th moment of the resistance, can by written as
\begin{eqnarray}
\langle \chi_+ (x ,x^\prime) R_+ (x ,x^\prime )^l \rangle_C / \langle \chi_+ (x ,x^\prime) \rangle_C \, . 
\end{eqnarray}

\subsection{Incorporation of noise}
In the usual RDN all the $\varsigma_{\underline{b}}$ are equal to one. Here, we are interested in a more general setup where the bond conductances fluctuate statically about some average value $\overline{\varsigma}$. In other words we are interested in an RDN with static noise. This noise is modeled by distributing the $\varsigma_{\underline{b}}$ according to some distribution function $f$ with mean $\overline{\varsigma}$ and higher cumulants $\Delta^{(l\geq 2)}$ satisfying $\Delta^{(l)} \ll \overline{\varsigma}^l$. The condition on the cumulants is imposed to avoid unphysical negative conductances. The distribution function $f$ might for example be Gaussian. Nevertheless, our considerations are not limited to this particular choice.

The noise average will be denoted by 
\begin{eqnarray}
\label{defOfNoiseAverage}
\{ \cdots \}_f = \int \prod_{\underline{b}} d \varsigma_{\underline{b}} \, f \left( \varsigma_b \right) \cdots
\end{eqnarray}
and its $l$th cumulant by 
\begin{eqnarray}
\{ \cdots^l \}^{(c)}_f = \left. \frac{\partial^l}{\partial 
\lambda^l} \ln \left\{ \exp \left[ \lambda \cdots \right] \right\}_f 
\right|_{\lambda =0}\, .
\end{eqnarray}
Of course, both kinds of disorder, the random dilution of the lattice and the noise, influence the statistical properties of the macroscopic resistance. Both are reflected mutually by the moments
\begin{eqnarray}
\label{defMoment}
M_R^{(l)}(x ,x^\prime) = \frac{\left\langle \chi_+ (x ,x^\prime) \left\{ R_+ (x ,x^\prime )^l \right\}_f \right\rangle_C}{\left\langle \chi_+ (x ,x^\prime) \right\rangle_C} 
\end{eqnarray}
and the cumulants 
\begin{eqnarray}
\label{defCumulant}
C_R^{(l)}(x ,x^\prime)
 = \frac{\big\langle \chi_+ (x ,x^\prime) \big\{ R_+ (x 
,x^\prime )^l \big\}_f^{(c)} \big\rangle_C}{\left\langle \chi_+ (x ,x^\prime) \right\rangle_C} \,  . 
\end{eqnarray}

\subsection{Moments of the current distribution}
\label{momentsOfTheCurrentDistribution}
Primarily we are interested in the moments of the current distribution defined by
\begin{eqnarray}
\label{multiMoment}
M_I^{(l)}(x ,x^\prime) = \left\langle \chi (x ,x^\prime) \sum_{\underline{b}} 
\left( \frac{I_{\underline{b}}}{I} \right)^{2l} \right\rangle_C / \left\langle \chi (x ,x^\prime) 
\right\rangle_C \, .
\nonumber \\ 
\end{eqnarray}
There exists an intimate relationship between the $M_I^{(l)}$ and the $C_R^{(l)}$ that can be understood as a consequence of Cohn's theorem~\cite{cohn_50}. We shall now briefly review this relationship.

In Eq.~(\ref{defOfNoiseAverage}) the noise average is defined as an average with respect to the distribution of the bond conductances $\sigma_{\underline{b}}$. Alternatively one might express the macroscopic resistance in terms of the bond resistances $\rho_{\underline{b}}$ and average over their distribution. Of course, not only the $\sigma_{\underline{b}}$ but also the $\rho_{\underline{b}}$ are distributed independently and equally. Assume that the distribution function of the deviations $\delta \rho_{\underline{b}} = \rho_{\underline{b}} - \overline{\rho}$ of the resistance of each bond from its average $\overline{\rho}$ has the form
\begin{eqnarray}
g_s \left( \delta \rho_{\underline{b}} \right) = \frac{1}{s} h \left( \frac{\delta \rho_{\underline{b}}}{s} \right)
\end{eqnarray}
and that
\begin{eqnarray}
\lim_{s\to 0} g_s \left( \delta \rho_{\underline{b}} \right) = \delta \left( \delta \rho_{\underline{b}} \right) \, .
\end{eqnarray}
$s$ is a variable with units of resistance which sets the scale of the distribution. With this form of $g_s$, the $n$th cumulant $v_n$ of $\delta \rho_b$ tends to zero as $s^n$. This follows from the generating function $c \left( \lambda s \right)$ of the $v_n$:
\begin{eqnarray}
\exp \left[ c \left( \lambda s \right) \right] &=& \left\{ \exp \left( \lambda \delta 
\rho_{\underline{b}} \right) \right\}_f = \int dy \ h \left( y \right) \exp \left( \lambda s y \right)
\nonumber \\ 
&=& \exp 
\left( \sum_{n=1}^\infty \frac{\lambda^n}{n!} v_n\right) \, ,
\end{eqnarray}
where $v_n = c_n s^n$ with $c_n$ being constants. In general $\left\{ R_+ (x ,x^\prime )^l \right\}_f^{(c)}$ depends on the entire set of cumulants $\left\{ v_n \right\}$. However, in the limit $s\to 0$ the leading term is proportional to $v_l$  as we will see in a moment. Consider the generating function $C \left( \lambda \right)$ of the cumulants 
$\left\{ R_+ (x ,x^\prime )^l \right\}_f^{(c)}$, 
\begin{eqnarray}
\exp \left[ C \left( \lambda \right) \right] = \int \prod_{\underline{b}} d \delta \rho_b \ g_s \left( 
\delta \rho_{\underline{b}} \right) \exp \left[ \lambda R (x ,x^\prime ) \right] \, .
\nonumber \\
\end{eqnarray}
Expansion of the macroscopic resistance in a power series in the $\delta \rho_{\underline{b}}$ leads to
\begin{eqnarray}
\label{schubidu}
&&\exp \left[ C \left( \lambda \right) \right] = \int \prod_{\underline{b}} d y_b \, h \left( y_b \right) 
\exp \Bigg[ \lambda R^0_+ (x ,x^\prime ) 
\nonumber \\
&+& \lambda \sum_{k=1}^\infty \sum_{b_1 , \cdots , 
b_k} \frac{s^k}{k!} \left. \frac{\partial^k R (x ,x^\prime )}{\partial \rho_{{\underline{b}}_1} \cdots 
\partial \rho_{{\underline{b}}_k}} \right|_{\overline{\rho}} y_{{\underline{b}}_1} \cdots y_{{\underline{b}}_k} \Bigg] \, ,
\nonumber \\
\end{eqnarray}
where $R^0_+ (x ,x^\prime )$ is the resistance when $\delta \rho_{\underline{b}} = 0$ for every bond ${\underline{b}}$. Equation~(\ref{schubidu}) can be rearranged as
\begin{eqnarray}
&&\exp \left[ C \left( \lambda \right) \right]
\nonumber \\
&=& \exp \Bigg[ \lambda R^0_+ (x ,x^\prime ) 
+ \lambda \sum_{k=2}^\infty \sum_{{\underline{b}}_1 , \cdots , {\underline{b}}_k} \frac{s^k}{k!}
 \frac{\partial^k R_+ (x ,x^\prime )}{\partial \rho_{{\underline{b}}_1} \cdots \partial \rho_{{\underline{b}}_k}} 
\bigg|_{\overline{\rho}} 
\nonumber \\
&\times&
\frac{\partial^k }{\partial z_{{\underline{b}}_1} \cdots \partial z_{{\underline{b}}_k}}
\Bigg] 
\prod_{\underline{b}} \exp \Big[ c \left( z_{\underline{b}} \right) \Big] \Bigg|_{\lambda s \sum_{\underline{b}} 
\left. 
\frac{\partial R_+ (x ,x^\prime )}{\partial \rho_{\underline{b}}} \right|_{\overline{\rho}}}
\nonumber \\
&=& \exp \Bigg[ \lambda R^0_+ (x ,x^\prime ) + \sum_{l=1}^\infty \left( \lambda s 
\right)^l 
c_l \sum_{\underline{b}} \left(  \left. \frac{\partial R_+ (x ,x^\prime )}{\partial \rho_b} 
\right|_{\overline{\rho}} \right)^l 
\nonumber \\
&+& \sum_{i=2}^\infty f_i \left( \lambda s^i \right)  
\Bigg] \, ,
\end{eqnarray}
where $f_i$ is a function of $\lambda s^i$. Hence, for $l\geq 2$,
\begin{eqnarray}
\left\{ R_+ (x ,x^\prime )^l \right\}_f^{(c)} = c_l \sum_{\underline{b}} \left( s \left. 
\frac{\partial R_+ (x ,x^\prime )}{\partial \rho_{\underline{b}}} \right|_{\overline{\rho}} \right)^l  
\Big( 1 + O ( s ) \Big) \, .
\nonumber \\
\end{eqnarray}
In the limit $s\to 0$ the leading term is
\begin{eqnarray}
\label{gleichFA}
\left\{ R_+ (x ,x^\prime )^l \right\}_f^{(c)} &=& v_l \sum_{\underline{b}} \left( \left. 
\frac{\partial R_+ (x ,x^\prime )}{\partial \rho_{\underline{b}}} \right|_{\overline{\rho}} \right)^l 
\nonumber \\
&=& v_l \sum_{\underline{b}} \left( \frac{I_{\underline{b}}}{I} \right)^{2l} \, ,
\end{eqnarray}
where we have used Cohn's Theorem in the form
\begin{eqnarray}
\label{originalCohn}
\frac{\partial R_+ (x ,x^\prime )}{\partial \rho_{\underline{b}}} = \left( \frac{I_{\underline{b}}}{I} \right)^2 \, .
\end{eqnarray}
Upon substitution of 
Eq.~(\ref{gleichFA}) into Eq.~(\ref{defCumulant}) one finds for the noise cumulants
\begin{eqnarray}
\label{finalCumulant}
C_R^{(l)}(x ,x^\prime) = v_l \, M_I^{(l)}(x ,x^\prime) \, , 
\end{eqnarray}
i.e., the $l$th noise cumulant is proportional to the $l$th multifractal moment of the current distribution. In the following we will exploit this relation in the sense that we will study the $C_R^{(l)}$ as a surrogate for the $M_I^{(l)}$. We will see below that the $C_R^{(l)}$ are accessible in an elegant way by renormalized field theory whereas we do not know of such an approach for the $M_I^{(l)}$ directly.

\subsection{Replication}
\label{replication}
In this section we devise a generating function for the $C_R^{(l)}$ based on the ideas of Stephen~\cite{stephen_78} and their refinement by Park, Harris, and Lubensky (PHL)~\cite{park_harris_lubensky_87}. We demonstrate that this generating function indeed serves its purpose and explain how the average conductance can be extracted from it.

PHL introduced the quantity
\begin{eqnarray}
\label{defPsi}
\psi_{\tens{\lambda}}(x) = \exp \left( i \tens{\lambda} \cdot \tens{V}_x \right) \, , \quad \tens{\lambda} \neq \tens{0} \, .
\end{eqnarray}
$\tens{V}_x$ is a $(D \times E)$-fold replicated variant of the voltage $V_x$ at lattice site $x$, 
\begin{eqnarray}
\tens{V}_x =
\left( 
\begin{array}{ccc}
V_x^{(1,1)} & \cdots & V_x^{(1,D)}\\
\vdots & \ddots & \vdots\\
V_x^{(E,1)} & \cdots & V_x^{(E,D)}
\end{array}
\right) \, .
\end{eqnarray}
$\tens{\lambda}$ is, apart from a factor $-i$, a replicated external current that is like $\tens{V}_x$ a matrix with $(D \times E)$ components. The corresponding scalar product is defined as $\tens{\lambda} \cdot \tens{V}_x = \sum_{\alpha , \beta =1}^{D,E} \lambda^{(\alpha ,\beta )} V_x^{(\alpha ,\beta )}$. 

In order to proceed towards the desired generating function we now consider the two-point correlation function of $\psi_{\tens{\lambda}}(x)$
\begin{eqnarray}
\label{def2Pkt}
G \left( x, x^\prime ,\tens{\lambda} \right) = \left\langle 
\psi_{\tens{\lambda}}(x)\psi_{-\tens{\lambda}}(x^\prime) 
\right\rangle_{\mbox{\scriptsize{rep}}}
\end{eqnarray}
where the average is defined by
\begin{eqnarray}
\label{devRepAve}
&&\Big\langle \cdots \Big\rangle_{\mbox{\scriptsize{rep}}} =
\Bigg\langle \Bigg\{ \frac{1}{\prod_{\beta =1}^E Z \left( \left\{ 
\sigma_{\underline{b}}^{(\beta )} \right\} , C \right)^D } 
\int \prod_j d \tens{V}_j
\nonumber \\
& & \times \, \exp \bigg[ -\frac{1}{2} P \left( \left\{ \vec{V}^{(\delta )} \right\} , \left\{ \sigma_b^{(\delta )} \right\} , C  \right) \bigg] \ \cdots \ \Bigg\}_f \Bigg\rangle_C \, .
\nonumber \\
\end{eqnarray}
Here, $d \tens{V}_j$ is an abbreviation for $\prod_{\alpha ,\beta=1}^{D,E} dV_j^{(\alpha , \beta )}$. 
\begin{eqnarray}
P \left( \left\{ \vec{V}^{(\delta )} \right\} , \left\{ \sigma_b^{(\delta )} \right\} , C  \right) 
=  \sum_{\gamma =1}^{D} \sum_{\underline{b}} \sigma_{\underline{b}}^{(\delta )} \left[ V_{\underline{b}}^{(\gamma ,\delta)} \right]^2
\nonumber \\
\end{eqnarray}
is the replicated version of the electric power with $\vec{V}_x^{(\delta )} = ( V_x^{(1 , \delta )} , \cdots , V_x^{(D , \delta )} )$. The normalization factor in Eq.~(\ref{devRepAve}) is given by
\begin{eqnarray}
\label{norm}
\label{noisyNorm}
&&Z \left( \left\{ \sigma_b^{(\beta )} \right\} , C \right) 
\nonumber \\
&&= \, \int \prod_{j} dV_{j} \exp 
\left[ -\frac{1}{2} P \left( \left\{ V \right\} , \left\{ \sigma_b^{(\beta )} \right\} , C \right) \right] \, .
\nonumber \\
\end{eqnarray}

At this point we would like to comment on regularization issues. First, it is important to realize that the integrands in Eqs.~(\ref{devRepAve}) and (\ref{norm}) depend only on voltage differences and hence the integrals are divergent. To give these integrals a well defined meaning one can introduce an additional power term $\frac{i\omega}{2} \sum_i V^2_i$. Physically the new term corresponds to grounding each lattice site by a capacitor of unit capacity. The original situation can be restored by taking the limit of vanishing frequency, $\omega \to 0$. Second, it is not guaranteed that $Z$ stays finite because infinite voltage drops may occur. Thus, the limit $\lim_{D \to 0}{Z^{DE}}$ is not well defined. This problem may be regularized by restricting the voltage variables to a finite interval. However, we have to bear in mind that $\tens{\lambda} = \tens{0}$ has to be excluded properly. We take care of both points simultaneously by resorting to a lattice regularization. To be specific, we switch to voltages taking discrete values on a $(D \times E)$-dimensional torus, henceforth called the replica space. In practice we set $\tens{\vartheta}= \Delta \vartheta \tens{k}$ with $\tens{k}$ being an $(D \times E)$-dimensional integer with $-M < k^{(\alpha ,\beta )} \leq M$ and $k^{(\alpha ,\beta)}=k^{(\alpha ,\beta )} \mbox{mod} (2M)$. $\Delta \vartheta = \vartheta_M /M$ is the gap between successive voltages and $\vartheta_M$ is the voltage cutoff. The continuum may be restored by taking $\theta_M \to \infty$ and $\Delta \theta \to 0$. By setting $M=m^2$, $\vartheta_M = \vartheta_0 m$, and, respectively, $\Delta \vartheta = \vartheta_0 /m$, the two limits can be taken simultaneously via $m \to \infty$. For the replica currents we set
\begin{eqnarray}
\tens{\lambda} = \Delta \lambda \tens{l} \ , \ \Delta \lambda \hspace{1mm} \Delta \theta = \pi / M \ ,
\end{eqnarray}
where $\tens{l}$ is a $(D \times E)$-dimensional integer taking the same values as $\tens{k}$. This choice guarantees that the completeness and orthogonality relations
\begin{subequations}
\label{noisyComplete}
\begin{eqnarray}
\frac{1}{(2M)^{DE}} \sum_{\tens{\vartheta}} \exp \left( 
i \tens{\lambda} \cdot \tens{\theta} \right) = \delta_{\tens{\lambda} ,\tens{0} 
\hspace{0.15em}\mbox{\scriptsize{mod}}(2M \Delta \lambda) }
\end{eqnarray}
and
\begin{eqnarray}
\frac{1}{(2M)^{DE}} \sum_{\tens{\lambda}} \exp 
\left( i \tens{\lambda} \cdot \tens{\theta} \right) = \delta_{\tens{\theta} ,\stackrel{\mbox{{\tiny $\leftrightarrow$}}}{0} 
\hspace{0.15em}\mbox{\scriptsize{mod}}(2M \Delta \theta)}
\end{eqnarray}
\end{subequations}
do hold. Equation~(\ref{noisyComplete}) provides us with a Fourier transform in replica space. It is important to note that the replica space Fourier transform of $\psi_{\tens{\lambda}}(x)$,
\begin{eqnarray}
\Phi_{\tens{\vartheta}} \left( x \right) &=& (2M)^{-DE} 
\sum_{\tens{\lambda} \neq \tens{0}} \exp \left( i \tens{\lambda} \cdot \tens{\vartheta} \right) \psi_{\tens{\lambda}} (x) 
\nonumber \\
&=& \delta_{\tens{\vartheta}, \tens{\vartheta}_{x}} - (2M)^{-DE} 
\end{eqnarray}
satisfies the condition $\sum_{\tens{\vartheta}} \Phi_{\tens{\vartheta}} ( x ) = 0$. Hence, it can be identified as a Potts spin~\cite{Zia_Wallace_75} with $q = (2M)^{DE}$ states. The relation of the RDN to the Potts model that emerges here as a consequence of the lattice regularization is important as well as intuitively plausible. That is because the purely geometric properties of the RDN are those of percolation, either IP or DP, depending on which sector of the phase diagram is considered. It is a well known fact that the Potts model describes percolation in the limit $q\to 1$~\cite{kasteleyn_fortuin_69}. Note that this limit corresponds to the replica limit via $q = (2M)^{DE}$.

Before proceeding with the evaluation of the correlation functions~(\ref{def2Pkt}) we would like to make one more comment on the replication procedure. The replication scheme employed here goes beyond the usual replica trick in the sense that it involves a second replication parameter, i.e., that the replicated quantities are $(D\times E)$-tuples and not just $D$-tuples. This subtlety has its origin in the definitions (\ref{defMoment}) and (\ref{defCumulant}) which require to treat the averages $\langle \cdots \rangle_C$ and $\left\{ \cdots \right\}_f$ independently. The great benefit of the replication is to provide for the free parameter $D$ which we may tune to zero~\cite{footnote1}. In this replica limit the normalization denominator $Z^{-DE}$ goes to one and hence does not depend on the distribution of the bond conductances anymore. Then the only remaining dependence on this distribution rests in the electric power $P$ appearing in the exponential in Eq.~(\ref{def2Pkt}). In the replica limit, therefore, we just have to average this exponential instead of the entire right hand side of Eq.~(\ref{def2Pkt}). This average then provides us with an effective electric power or Hamiltonian which serves as vantage point for all further calculations. The effective Hamiltonian will be discussed in Sec.~\ref{fieldTheoreticHamiltonian}.

Now we come back to the role of the correlation functions~(\ref{def2Pkt}) as a generating function. The integrations associated with the average~(\ref{devRepAve}) are not Gaussian. However, they can be carried out in an approximating manner by applying the saddle point method as it was done by Harris~\cite{harris_87} in the related problem of nonlinear random resistor networks. We extract the leading contribution to the integral stemming from the maximum of the integrand. This maximum is determined by the solution of Kirchhoff's equations, i.e., by the macroscopic resistance. The conditions under which the saddle point method works reliable will be outlined in the next paragraph. Under these conditions we find 
\begin{eqnarray}
\label{noisyGenFkt}
G \left( x, x^\prime ,\tens{\lambda} \right) 
= 
\left\langle \prod_{\beta =1}^E \left\{ \exp \left[ - 
\frac{\vec{\lambda}^{(\beta )2}}{2} R_+^{(\beta )} \left( x,x^\prime 
\right) \right] \right\}_f \right\rangle_C \, ,
\nonumber \\
\end{eqnarray}
up to an unimportant multiplicative constant that goes to one in the replica limit $D \to 0$. On defining $K_l ( \tens{\lambda} ) = \sum_{\beta =1}^E [ \sum_{\alpha =1}^D ( \lambda^{(\alpha ,\beta )} )^2 ]^l$ we obtain by expanding in terms of cumulants
\begin{eqnarray}
\label{cumulantGenFkt1}
&&G \left( x, x^\prime ,\tens{\lambda} \right) =
\nonumber \\
&& 
\left\langle  \exp \left[ \sum_{l=1}^\infty \frac{(-1/2)^l}{l!} K_l \left( \tens{\lambda} \right) \left\{ R_+ 
\left( x,x^\prime \right)^l \right\}_f^{(c)} \right]  \right\rangle_C \, .
\nonumber \\
\end{eqnarray}
Upon expanding the exponential we get
\begin{eqnarray}
\label{cumulantGenFkt2}
&&G \left( x, x^\prime ,\tens{\lambda} \right) = \left\langle \chi_+ ( x, x^\prime) \right\rangle_C
\nonumber \\
&& 
\times \, \Bigg\{ 1 + \sum_{l=1}^\infty \frac{(-1/2)^l}{l!} K_l \left( \tens{\lambda} \right) C_R^{(l)} ( x, x^\prime) + \cdots \Bigg\} .
\nonumber \\
\end{eqnarray}
 We learn here that $C_R^{(l)}$ can be calculated via
\begin{eqnarray}
\label{exploitGenFkt}
&&C_R^{(l)} \left( x, x^\prime \right) = \left\langle \chi_+ ( x, x^\prime) \right\rangle_C^{-1}
\nonumber \\
&&
\times \,
\frac{\partial}{\partial \left[ \frac{(-1/2)^l}{l!} K_l \left( \tens{\lambda} \right) \right]}\,  G \left( x, x^\prime ,\tens{\lambda} \right) 
\bigg|_{\tens{\lambda} = \tens{0}} \, ,
\nonumber \\
\end{eqnarray}
i.e., $G$ represents indeed the desired generating function for the noise cumulants.

Now to the conditions for the saddle point approximation.  We work near the limit when all the components of $\tens{\lambda}$ are equal and continue to large imaginary values. Accordingly we set~\cite{harris_87}
\begin{eqnarray}
\label{lambdaChoice}
\lambda^{(\alpha , \beta)} = i \lambda_0 + \xi^{(\alpha , \beta )}
\end{eqnarray}
with real $\lambda_0$ and $\xi^{(\alpha , \beta)}$, $\sum_{\alpha =1}^D \xi^{(\alpha , \beta )} = 0$. The saddle point approximation gives Eq.~(\ref{noisyGenFkt}) provided that
\begin{eqnarray}
\label{cond1}
\left| \lambda_0 \right| \gg 1 \, . 
\end{eqnarray}
On the other hand one has
\begin{eqnarray}
 \vec{\lambda}^{(\beta )2} = - D  \lambda_0^2 + \vec{\xi}^{(\beta )2} \, .
\end{eqnarray}
Thus, one can justify the expansion in Eq.~(\ref{cumulantGenFkt2}) by invoking the conditions
\begin{eqnarray}
\label{cond2}
\lambda_0^{2} \ll D^{-1} \quad \mbox{and} \quad \vec{\xi}^{(\beta )2} \ll 1 \, .
\end{eqnarray}
It is important to realize that the replica limit $D\to 0$ allows for a simultaneous fulfillment of the conditions (\ref{cond1}) and (\ref{cond2}).

\subsection{Field theoretic Hamiltonian}
\label{fieldTheoreticHamiltonian}
This section presents the derivation of our field theoretic Hamiltonian for the noisy RDN. We start by revisiting Eq.~(\ref{devRepAve}) and note that the effective Hamiltonian heralded in Sec.~\ref{replication} is given by
\begin{eqnarray}
\label{effHamil1}
H_{\mbox{\scriptsize{rep}}} &=&  - \ln \left\langle  \exp \left[ - \frac{1}{2} P \left( \left\{ \vec{\vartheta} \right\} \right) \right] \right\rangle_C \, .
\end{eqnarray}
Carrying out the average over the diluted lattice configurations leads to
\begin{eqnarray}
\label{effHamil2}
H_{\mbox{\scriptsize{rep}}} = - \sum_{\underline{b}} K \left( \vec{\vartheta}_{\underline{b}} \right) \, ,
\end{eqnarray}
where we have introduced
\begin{eqnarray}
\label{kern1}
&&K \left( \vec{\vartheta}\right) = \ln \Bigg\{ q + p \prod_{\beta =1}^E \bigg\{ \exp \left[ - \frac{\varsigma^{(\beta )}}{2} \vec{\vartheta}^{(\beta )2} \right] \bigg\}_f
\nonumber \\
&+& p_+ \prod_{\beta =1}^E \bigg\{ \prod_{\alpha =1}^D \exp \left[ - \frac{\varsigma^{(\beta )}}{2} \theta \left( \vartheta^{(\alpha ,\beta )} \right) \vartheta^{(\alpha , \beta )2} \right] \bigg\}_f
\nonumber \\
&+& p_- \prod_{\beta =1}^E \bigg\{ \prod_{\alpha =1}^D \exp \left[ - \frac{\varsigma^{(\beta )}}{2} \theta \left( - \vartheta^{(\alpha ,\beta )} \right) \vartheta^{(\alpha ,\beta )2} \right]  \bigg\}_f \Bigg\} \, .
\nonumber \\
\end{eqnarray}
Now recall our choice for $\tens{\lambda}$ in Eq.~(\ref{lambdaChoice}). Since $\tens{\lambda}$ and $\tens{\vartheta}$ are related via Ohm's law Eq.~(\ref{avoid}), we have to choose $\tens{\vartheta}$ consistently:
\begin{eqnarray}
\vartheta^{(\alpha ,\beta)} = \vartheta_0 + \zeta^{(\alpha ,\beta)}
\end{eqnarray}
with real $\vartheta_0$ and $\zeta^{(\alpha ,\beta)}$, $\sum_{\alpha =1}^D \zeta^{(\alpha ,\beta)} = 0$. We impose the conditions
\begin{eqnarray}
\left| \vartheta_0 \right| \gg 1 \ , \quad \vartheta_0^2 \ll D^{-1} \ , \quad \vec{\zeta}^{(\beta )2} \ll 1 \, .
\end{eqnarray}
Under these conditions we have $\theta ( \vartheta^{(\alpha ,\beta )} ) = \theta ( \vartheta_0 )$. Hence, we can write
\begin{eqnarray}
\label{kern2}
&&K \left( \vec{\vartheta}\right) = \ln \Bigg\{ q + p_+ \theta \left( -  \vartheta_0 \right) + p_-  \theta \left( \vartheta_0 \right)
\nonumber \\
&+& \big[ p + p_+ \theta \left( \vartheta_0 \right) + p_-  \theta \left( - \vartheta_0 \right) \big] 
\nonumber \\
&\times&
\exp \left[ \sum_{l=1}^\infty \frac{(-1/2)^l}{l!} 
\Delta^{(l)} K_l \left( \tens{\vartheta} \right) 
\right] \Bigg\} \, .
\end{eqnarray}
After some additional algebraic steps and by dropping a term
\begin{eqnarray}
\theta \left( \vartheta_0 \right) \ln \left[ 1 - p - p_+ \right] + \theta \left( - \vartheta_0 \right) \ln \left[ 1 - p - p_- \right] \, ,
\end{eqnarray}
that does not depend on the bond conductances we arrive at
\begin{eqnarray}
\label{kern3}
K \left( \vec{\vartheta}\right) = \theta \left( \vartheta_0 \right) K_+ \left( \vec{\vartheta}\right) + \theta \left( - \vartheta_0 \right) K_- \left( \vec{\vartheta}\right) \, ,
\end{eqnarray}
with
\begin{eqnarray}
\label{defKpm}
&&K_\pm \left( \vec{\vartheta}\right) = 
\nonumber \\
&&\ln \left\{ 1 + \frac{p + p_\pm}{ 1 - p - p_\pm} \exp \left[ \sum_{l=1}^\infty \frac{(-1/2)^l}{l!} 
\Delta^{(l)} K_l \left( \tens{\vartheta} \right) \right] \right\} \, .
\nonumber \\
\end{eqnarray}
In order to proceed towards a field theoretic Hamiltonian we expand $K ( \tens{\vartheta} )$ in terms of the $\psi_{\tens{\lambda}}(i) = \exp ( i \tens{\lambda} \cdot \tens{\vartheta}_i )$. Recalling that $\underline{b}$ is a shorthand for the bond $\underline{b}_{<i,j>}$ between sites $i$ and $j$ we write
\begin{eqnarray}
\label{kern4}
K \left( \tens{\vartheta}_{\underline{b}} \right) &=& \frac{1}{\left( 2M \right)^{DE}} \sum_{\tens{\lambda}} \sum_{\tens{\vartheta}} \exp \left[ i \tens{\lambda} \cdot \left( \tens{\vartheta}_{\underline{b}} - \tens{\vartheta} \right) \right] K \left( \tens{\vartheta} \right)
\nonumber \\
&=& \sum_{\tens{\lambda} \neq \vec{0}} \psi_{\tens{\lambda}} \left( i \right) \psi_{-\tens{\lambda}} \left( j \right) 
\nonumber \\
&\times& \left[  \theta \left( \lambda_0 \right)\widetilde{K}_+ \left( \tens{\lambda} \right) + \theta \left( -\lambda_0 \right)\widetilde{K}_- \left( \tens{\lambda} \right) \right] \, .
\end{eqnarray}
Here, we have exploited that $\theta ( \vartheta_0 ) = \theta ( \lambda_0 )$. $\widetilde{K}_\pm ( \tens{\lambda})$ stands for the Fourier transform of $K_\pm ( \tens{\vartheta} )$, 
\begin{eqnarray}
\label{rrr}
\widetilde{K}_\pm \left( \tens{\lambda} \right) = \frac{1}{\left( 2M \right)^{DE}} \sum_{\tens{\vartheta}} \exp \left[ i \tens{\lambda} \cdot \tens{\vartheta} \right] K_\pm \left( \tens{\vartheta} \right) \, .
\end{eqnarray}
For evaluating Eq.~(\ref{rrr}) we switch back to continuous voltages,
\begin{eqnarray}
&&\widetilde{K}_\pm \left( \tens{\lambda} \right) = 
\int_{-\infty}^\infty d \tens{\vartheta} \ \exp \left( 
-i \tens{\lambda} \cdot \tens{\vartheta} \right) \ln \Bigg\{ 1 + \frac{p + p_\pm}{1-p-p_\pm} 
\nonumber \\
&& \times \, \exp \left[ \sum_{l=1}^\infty \frac{(-1/2)^l}{l!} \Delta^{(l)} K_l \left( \tens{\vartheta} \right) \right] \Bigg\} \, , 
\end{eqnarray}
where we have dropped a factor $(2\vartheta_M )^{-DE}$. Taylor expansion of the logarithm yields a series of terms of the form
\begin{eqnarray}
\label{s1} 
\int_{-\infty}^\infty d \tens{\vartheta} \ \exp \left[ -i \tens{\lambda} \cdot 
\tens{\vartheta} - a \overline{\varsigma} \tens{\vartheta}^2 - \sum_{l=2} b_l \left( 
\overline{\varsigma} s \right)^l K_l \left( \tens{\vartheta} 
\right) \right]  \, ,
\nonumber \\ 
\end{eqnarray}
where $a$ the $b_l$ are constants of order $O \left( s^0 \right)$. For notational simplicity we dropped the subscript $\pm$. In addition to the expansion of the logarithm we expand in a power series in $s$, 
\begin{eqnarray}
\label{s2} 
\mbox{Eq.~(\ref{s1})} &=& \int_{-\infty}^\infty d \tens{\vartheta} \ \exp \left[ -i \tens{\lambda} \cdot 
\tens{\vartheta} - a \overline{\varsigma} 
\tens{\vartheta}^2 \right] 
\nonumber \\
&\times& \left\{ 1 + \sum_{l=2}^\infty \left( \overline{\varsigma} s \right)^l P_l \left( \tens{\vartheta} \right) \right\} \, . 
\end{eqnarray}
Here, the $P_l$ are homogeneous polynomials of order $2l$ in $\tens{\lambda}$ which are a sums of terms proportional to
\begin{eqnarray}
\prod_{i \geq 2} K_i \left( \tens{\vartheta} \right)^{l_i} 
\end{eqnarray} 
such that $\sum_i i l_i = l$. Completing squares in the exponential in Eq.~(\ref{s2}) gives
\begin{eqnarray}
\label{s3} 
\mbox{Eq.~(\ref{s2})} &=& \exp \left[ - \frac{\tens{\lambda}^2}{4 a \overline{\varsigma}} \right] \int_{-\infty}^\infty d 
\tens{\vartheta} \ \exp \left[ - a \overline{\varsigma} 
\tens{\vartheta}^2 \right] 
\nonumber \\
&\times&
\left\{ 1 + 
\sum_{l=2}^\infty 
\left( \overline{\varsigma} s \right)^l P_l \left( \tens{\vartheta} - i \frac{\tens{\lambda}}{2 a \overline{\varsigma}} \right) \right\} 
\nonumber \\
&=& \exp \left[ - \frac{\tens{\lambda}^2}{4 a \overline{\varsigma}} \right] \Bigg\{ 1 + \sum_{l=2}^\infty \left( \overline{\varsigma} s 
\right)^l \bigg[ P_l \left( \frac{\tens{\lambda}}{ 
\overline{\varsigma}} \right) + \cdots 
\nonumber \\
&+& \overline{\varsigma}^{(-r)} P_{l-r} \left( \frac{\tens{\lambda}}{ \overline{\varsigma}} \right) + 
\cdots \bigg] \Bigg\} \, , 
\end{eqnarray}
where we have omitted multiplicative factors decorating the $P_l$. Due to the homogeneity of the $P_l$, Eq.~(\ref{s3}) can be rearranged as
\begin{eqnarray}
\label{s4} 
&& \mbox{Eq.~(\ref{s3})} 
\nonumber \\
& & = \, \exp \left[ - \frac{\tens{\lambda}^2}{4 a \overline{\varsigma}} \right] \Bigg\{ 1 + \sum_{l=2}^\infty  s^l \bigg[ \overline{\varsigma}^{-l} P_l \left( \tens{\lambda} \right) + \cdots 
\nonumber \\
&& +\, \overline{\varsigma}^{-(l-r)}P_{l-r} \left( \tens{\lambda} \right) + \cdots \bigg] \Bigg\} 
\nonumber \\
&& = \,  \exp \left[ - \frac{\tens{\lambda}^2}{4 a \overline{\varsigma}} \right] \left\{ 1 + \sum_{l^\prime =1}^\infty  \left( \frac{s}{\overline{\varsigma}} \right)^{l^\prime} \bigg[ 1 + O \left( s \right) 
\bigg] P_{l^\prime} \left( \tens{\lambda} \right) \right\} 
\, ,
 \nonumber \\
\end{eqnarray}
up to multiplicative factors. By keeping only the leading contributions, we deduce for $\widetilde{K} ( \tens{\lambda} )$ that
\begin{eqnarray}
\label{fullNoisyTaylorExp}
\widetilde{K}_\pm \left( \tens{\lambda} \right) = \tau_\pm + \sum_{p=1}^\infty w_{\pm ,p} \tens{\lambda}^{2p} + \sum_{P_l} v_{\pm ,P_l} P_l \left( \tens{\lambda} 
\right) \, ,
\nonumber \\
\end{eqnarray}
with $\tau_\pm$, $w_{\pm ,p} \sim \overline{\varsigma}^{-p}$, and $v_{\pm ,P_l} \sim \Delta^{(l)} / \overline{\varsigma}^{2l}$ being expansion coefficients.

The terms $w_p \tens{\lambda}^{2p}$ are irrelevant in the renormalization group sense for $p \geq 2$, cf.\ Sec.~\ref{relevance}. The $v_{P_l} P_l ( \tens{\lambda} )$ are irrelevant as well. However, we will demonstrate in Sec.~\ref{relevance} that the terms proportional to $K_l ( \tens{\lambda} )$ are indispensable in studying the noise cumulants; they are dangerously irrelevant. Therefore, we restrict the expansion of $\widetilde{K}_\pm ( \tens{\lambda} )$ to
\begin{eqnarray}
\label{noisyTaylorExp}
\widetilde{K}_\pm \left( \tens{\lambda} \right) = \tau_\pm + w_\pm \tens{\lambda}^2 + \sum_{l=2}^{\infty} v_{\pm ,l} K_l 
\left( \tens{\lambda} \right) \, ,
\end{eqnarray}
with $w_\pm = w_{\pm ,1}$, and $v_{\pm ,l} = v_{\pm ,K_l}$. Nevertheless, the neglected terms will regain some importance later on since they are required for the renormalization of the $v_{\pm ,l}$. The expansion coefficients in Eq.~(\ref{noisyTaylorExp}) satisfy $\tau_\pm (p, p_+ ,p_-) = \tau_\mp (p, p_- ,p_+)$, $w_\pm (p, p_+ ,p_-) = w_\mp (p, p_- ,p_+)$, and $v_{\pm ,l} (p, p_+ ,p_-) = v_{\mp ,l} (p, p_- ,p_+)$. The $K_l$ are homogeneous polynomials of order $2l$. For $l\geq 2$ they are possessing a $S [ O  (D )^E ]$ symmetry whereas $\tens{\lambda}^2$ has a full $O(DE)$ rotation symmetry in replica space.

Next we decompose $K ( \tens{\vartheta}_{\underline{b}} )$ into two parts, one being even and the other being odd under $\vec{\lambda} \to -\vec{\lambda}$:
\begin{eqnarray}
\label{kern5}
K \left( \tens{\vartheta}_{\underline{b}} \right)
&=& \sum_{\tens{\lambda} \neq \vec{0}} \psi_{\tens{\lambda}} \left( i \right) \psi_{-\tens{\lambda}} \left( j \right) 
\bigg\{ \frac{1}{2} \left[ \widetilde{K}_+ \left( \tens{\lambda} \right) + \widetilde{K}_- \left( \tens{\lambda} \right) \right] 
\nonumber \\
&+& \frac{1}{2} \left[ \theta \left( \lambda_0 \right) - \theta \left( -\lambda_0 \right) \right] \left[ \widetilde{K}_+ \left( \tens{\lambda} \right) - \widetilde{K}_- \left( \tens{\lambda} \right) \right] \bigg\} \, .
\nonumber \\
\end{eqnarray}
Then we insert Eq.~(\ref{kern5}) into Eq.~(\ref{effHamil2}). We also carry out a gradient expansion in position space which is justified because the interaction is short ranged. We find 
\begin{eqnarray}
\label{effHamil3}
H_{\mbox{\scriptsize{rep}}} &=&  - \sum_{\tens{\lambda} \neq \tens{0}} \sum_{i , \underline{b}_i} \bigg\{ \frac{1}{2} \left[ \widetilde{K}_+ \left( \tens{\lambda} \right) + \widetilde{K}_- \left( \tens{\lambda} \right) \right] 
\nonumber \\
&\times&
\psi_{-\tens{\lambda}} \left( i \right) \left[ 1  + \frac{1}{2} \left( \underline{b}_i \cdot \nabla \right)^2 + \cdots \right] \psi_{\tens{\lambda}} \left( i \right)
\nonumber \\
&+&  \frac{1}{2} \left[ \theta \left( \lambda_0 \right) - \theta \left( -\lambda_0 \right) \right] \left[ \widetilde{K}_+ \left( \tens{\lambda} \right) - \widetilde{K}_- \left( \tens{\lambda} \right) \right] 
\nonumber \\
&\times&
\psi_{-\tens{\lambda}} \left( i \right) \left[ \underline{b}_i \cdot \nabla + \cdots \right] \psi_{\tens{\lambda}} \left( i \right) \, .
\end{eqnarray}

We proceed with the usual coarse graining step and replace the 
$\psi_{\tens{\lambda}} ( i )$ by order parameter fields $\psi_{\tens{\lambda}} ( {\rm{\bf x}})$ which inherit the constraint $\tens{\lambda} \neq \tens{0}$. We model the corresponding field theoretic Hamiltonian $\mathcal{H}$ in the spirit of Landau as a mesoscopic free energy and introduce the Landau-Ginzburg-Wilson type functional
\begin{subequations}
\label{totalHamiltonian}
\begin{eqnarray}
\label{hamiltonian}
{\mathcal{H}} &=&  \int d^dx \Bigg\{ \frac{1}{2} \sum_{\tens{\lambda} \neq \tens{0}} \psi_{-\tens{\lambda}} \left( {\rm{\bf x}} \right)K \left( \nabla , \tens{\lambda} \right)  \psi_{\tens{\lambda}} \left( {\rm{\bf x}} \right)
\nonumber \\
&+&  \frac{g}{6} \sum_{\tens{\lambda}, \tens{\lambda}^\prime  , \tens{\lambda} + \tens{\lambda}^\prime \neq \tens{0}} \psi_{-\tens{\lambda}} \left( {\rm{\bf x}} \right) \psi_{-\tens{\lambda}^\prime} \left( {\rm{\bf x}} \right) \psi_{\tens{\lambda} + \tens{\lambda}^\prime} \left( {\rm{\bf x}} \right) \Bigg\} \, ,
\nonumber \\
\end{eqnarray}
where 
\begin{eqnarray}
K \left( \nabla , \tens{\lambda} \right) &=& \tau - \nabla^2 + w \tens{\lambda}^2 + \sum_{l=2}^{\infty} v_l K_l \left( \tens{\lambda} \right)
\nonumber \\
&+& \left[ \theta \left( \lambda_0 \right) - \theta \left( -\lambda_0 \right) \right] {\rm{\bf r}} \cdot \nabla   \, .
\end{eqnarray}
\end{subequations} 
In Eq.~(\ref{totalHamiltonian}) we have discarded terms of higher order in the fields that are irrelevant in the renormalization group sense. The parameter $\tau$ is the coarse grained relative of $\tau_+ + \tau_-$. It specifies the ``distance'' from the critical surface under consideration. In mean filed theory the transition occurs at $\tau =0$. $w \sim \sigma^{-1}$ is the coarse grained analog of $w_+ + w_-$. The $v_l$ stem from $v_{+,l} + v_{-,l}$. The vector ${\rm{\bf r}}$ lies in the preferred direction, ${\rm{\bf r}} = r {\rm{\bf n}}$. $\tau$, $w$, $v_l$, and $r$ depend on the three probabilities $p$, $p_+$, and $p_-$. $r$ is zero if $p_+ = p_-$. We will see as we go along that our Hamiltonian ${\mathcal{H}}$ describes in the limits $w\to 0$ and $v_l \to 0$ the usual purely geometric DP. Indeed ${\mathcal{H}}$ leads for $w\to 0$ and $v_l \to 0$ to exactly the same perturbation series as obtained in~\cite{cardy_sugar_80,janssen_81,janssen_2000}.

\subsection{Relevance}
\label{relevance}
In this section we show that the $v_l$ are dangerously irrelevant. The notion of {\sl dangerously irrelevant variables} was coined by Fisher~\cite{fisher_74}. It applies to variables that cannot be taken to zero because the quantity examined either vanishes or diverges in this limit. Later on the notion was carried over to field theory by Amit and Peliti~\cite{amit_peliti_82}. A characteristic feature of dangerously irrelevant variables is that corrections due to them determine the asymptotic behavior of quantities with the above property, so that their effect is felt arbitrarily close to a phase transition~\cite{diehl_86}. In contrast, usual irrelevant variables cause corrections to scaling that vanish at criticality.

We will see in a moment that the $v_l$ are irrelevant on dimensional grounds, i.e., they are associated with a negative naive dimension. However, we cannot simply take the $v_l$ to zero by appealing to their irrelevance, because the amplitudes of the noise cumulants vanish in this limit.

Now we embark on a scaling analysis in the current variable by rescaling $\tens{\lambda} \to b^{-1} \tens{\lambda}$. Here $b$ denotes a scaling factor and should not be confused with the index labeling the bonds. By substituting $\psi_{\tens{\lambda}} ( {\rm{\bf x}} ) = \psi_{b^{-1} \tens{\lambda}}^\dag ( {\rm{\bf x}} )$ into the Hamiltonian we get
\begin{eqnarray}
\label{scaling1}
&&\mathcal{H} \left[ \psi_{b^{-1} \tens{\lambda}}^\dag ( {\rm{\bf x}} ) ; \tau, r, ,w, \{ v_l \} \right] 
\nonumber \\
&& = \, \int d^dx \, \Bigg\{ \frac{1}{2} \sum_{\tens{\lambda} \neq \tens{0} } \psi_{b^{-1} \tens{\lambda}}^\dag ( {\rm{\bf x}} ) \, K \left( \nabla 
,\tens{\lambda} \right) \psi_{- b^{-1} \tens{\lambda}}^\dag ( {\rm{\bf x}} )
\nonumber \\
&& + \, 
\frac{g}{6} \sum_{\tens{\lambda}, \tens{\lambda}^\prime  , \tens{\lambda} + \tens{\lambda}^\prime \neq \tens{0}} \psi_{-b^{-1}\tens{\lambda}}^\dag \left( {\rm{\bf x}} \right) \psi_{-b^{-1}\tens{\lambda}^\prime}^\dag \left( {\rm{\bf x}} \right) 
\nonumber \\
&& \times \,
\psi_{b^{-1}\tens{\lambda} + b^{-1}\tens{\lambda}^\prime}^\dag \left( {\rm{\bf x}} \right) \Bigg\} \, .
\end{eqnarray}
Renaming the scaled voltage variables $\tens{\lambda}^\dag = b^{-1} \tens{\lambda}$ leads to
\begin{eqnarray}
\label{scaling2}
&&\mathcal{H} \left[ \psi_{\tens{\lambda}^\dag}^\dag ( {\rm{\bf x}} ) ; \tau, r, ,w, \{ v_l \} \right] 
\nonumber \\
&& = \, \int d^dx \, \Bigg\{ \frac{1}{2} \sum_{\tens{\lambda} \neq \tens{0} } \psi_{\tens{\lambda}^\dag}^\dag ( {\rm{\bf x}} ) \, K \left( \nabla 
, b \tens{\lambda}^\dag \right) \psi_{-\tens{\lambda}^\ast}^\dag ( {\rm{\bf x}} )
\nonumber \\
&& + \, 
\frac{g}{6} \sum_{\tens{\lambda}, \tens{\lambda}^\prime  , \tens{\lambda} + \tens{\lambda}^\prime \neq \tens{0}} \psi_{-\tens{\lambda}^\dag}^\dag \left( {\rm{\bf x}} \right) \psi_{-\tens{\lambda}^{\prime ^\dag}}^\dag \left( {\rm{\bf x}} \right) 
\psi_{\tens{\lambda}^\dag +\tens{\lambda}^{\prime ^\dag}}^\dag \left( {\rm{\bf x}} \right) \Bigg\} \, .
\nonumber \\
\end{eqnarray}
Obviously, a scaling of the current variable results in a scaling of the current cutoff $\lambda_M = \Delta \lambda \, M = \pi m/\vartheta_0$, {\em viz.} $\lambda_M \to b^{-1} \lambda_M$. However, by taking the limit $D \to 0$ and then $m \to \infty$, the dependence of the theory on the cutoff drops out. In other words: $\lambda_M$ is a redundant scaling variable. Thus, one can identify $\tens{\lambda}^\dag$ and $\tens{\lambda}$ which leads to the conclusion that 
\begin{eqnarray}
\label{relForH}
&&\mathcal{H} \left[ \psi_{b^{-1} \tens{\lambda}} ( {\rm{\bf x}} ) ; \tau ,r,w, \{ v_l \} \right] 
\nonumber \\
&&=\, \mathcal{H} \left[ \psi_{\tens{\lambda}} ( {\rm{\bf x}} ) ; \tau ,r, b^2 w, \{ b^{2l} v_l \} \right] \, . 
\end{eqnarray}

Now we consider the consequences of Eq.~(\ref{relForH}) for the correlation functions of the field $\psi_{\tens{\lambda}} ( {\rm{\bf x}} )$ defined by
\begin{eqnarray}
\label{correl}
&&G_N \left( \left\{ {\rm{\bf x}} , \tens{\lambda} \right\} ; \tau ,r, w,\{ v_l \} \right) 
\nonumber \\
&& =\, \int \mathcal{D} \psi \ \psi_{\tens{\lambda}_1} ( {\rm{\bf x}}_1 ) \cdots \psi_{\tens{\lambda}_N} ( {\rm{\bf x}}_N ) 
\nonumber \\
& & \times \exp \left( - \mathcal{H} \left[ \psi_{\tens{\lambda}} ( {\rm{\bf x}} ) ; \tau ,r, w, \{ v_l \} \right] \right) \, ,
\end{eqnarray}
where $\mathcal{D} \psi$ indicates an integration over the set of variables $\{ \psi_{\tens{\lambda}} ( {\rm{\bf x}} ) \}$ for all ${\rm{\bf x}}$ and $\tens{\lambda}$. Equation~(\ref{relForH}) implies that
\begin{eqnarray}
\label{ohnasch}
&&G_N \left( \left\{ {\rm{\bf x}} ,\tens{\lambda} \right\} ; \tau ,r , w, \{ v_l \}  \right) 
\nonumber \\
&&=\, G_N \left( \left\{ {\rm{\bf x}} , b^{-1} \tens{\lambda} \right\} ; \tau ,r, b^2 w, \{ b^{2l} v_l \} \right) \, .
\end{eqnarray}
From Eq.~(\ref{ohnasch}) in conjunction with Eq.~(\ref{cumulantGenFkt2}) we deduce
\begin{eqnarray}
&&K_l \left( \tens{\lambda} \right) C_R^{(l)} \left( \left( {\rm{\bf x}}, {\rm{\bf x}}^\prime \right) ; \tau ,r, w, \left\{ v_k \right\} \right)
\nonumber \\
&&= \, b^{-2l} K_l \left( \tens{\lambda} \right) C_R^{(l)} \left( \left( {\rm{\bf x}}, {\rm{\bf x}}^\prime \right) ; \tau ,r, b^2 w, \left\{ b^{2k} v_k \right\} \right) \, .
\nonumber \\
\end{eqnarray}
We are free to choose $b^2 = w^{-1}$. This choice gives
\begin{eqnarray}
\label{cumulantScaling}
&& C_R^{(l)} \left( \left( {\rm{\bf x}}, {\rm{\bf x}}^\prime \right) ; \tau ,r, w, \left\{ v_k \right\} \right) 
\nonumber \\
&&= \, w^l f_l \left( \left( {\rm{\bf x}}, {\rm{\bf x}}^\prime \right) ; \tau ,r, \left\{ \frac{v_k}{w^k} \right\} \right) \, ,
\end{eqnarray}
where $f_l$ is a scaling function. We learn from Eq.~(\ref{cumulantScaling}) that the coupling constants $v_k$ appear only in the combination $v_k / w^k$. A trivial consequence of the fact that the Hamiltonian~(\ref{totalHamiltonian}) must be dimensionless is that $w \tens{\lambda}^2 \sim \mu^2$ and  $v_k K_k ( \tens{\lambda} ) \sim \mu^2$, where $\mu$ is an inverse length scale. In other words $w \tens{\lambda}^2$ and $v_k K_k ( \tens{\lambda} )$ have a naive dimension 2. Thus, $v_k / w^k \sim \mu^{2-2k}$ and hence the $v_k / w^k$ have a negative naive dimension. This leads to the conclusion that the $v_{k}$ are irrelevant couplings.

Though irrelevant, one must not set $v_{l} =0$ in calculating the noise exponents. In order to see this we expand the scaling function $f_l$ in Eq.~(\ref{cumulantScaling}),
\begin{eqnarray}
\label{expOfCumulantScaling1}
&&C_R^{(l)} \left( \left( {\rm{\bf x}}, {\rm{\bf x}}^\prime \right) ; \tau ,r, w, \left\{ v_k \right\} \right) 
\nonumber \\
&& \, = w^l \left\{ C_l^{(l)} \frac{v_l}{w^l} + C_{l+1}^{(l)} 
\frac{v_{l+1}}{w^{l+1}} + \cdots \right\} \, ,
\end{eqnarray}
with $C_k^{(l)}$ being expansion coefficients depending on ${\rm{\bf x}}$, ${\rm{\bf x}}^\prime$, $\tau$, and $r$. It is important to recognize that $C_{k<l}^{(l)} = 0$ because the corresponding terms are not generated in the perturbation calculation. Equation~(\ref{expOfCumulantScaling1}) can be rewritten as
\begin{eqnarray}
\label{expOfCumulantScaling2}
&&C_R^{(l)} \left( \left( {\rm{\bf x}}, {\rm{\bf x}}^\prime \right) ; \tau ,r , w, \left\{ v_k \right\} \right) 
\nonumber \\
&&= \, v_l \left\{ C_l^{(l)} + C_{l+1}^{(l)} \frac{v_{l+1}}{w v_l} + \cdots \right\} \, .
\end{eqnarray}
Note that $v_{l+k}/(w^k v_l) \sim \mu^{-2k}$, i.e., the corresponding terms are irrelevant. The first term on the right hand side gives the leading behavior. Thus, $C_R^{(l)}$ vanishes upon setting $v_l = 0$ and we cannot gain any further information about $C_R^{(l)}$. In particular we cannot determine the associated noise exponent. In other words, the $v_{l}$ are dangerously irrelevant in investigating the critical properties of the $C_R^{(l\geq 2)}$.

We conclude this section by demonstrating that the $w_{p\geq 2}$ appearing in Eq.~(\ref{fullNoisyTaylorExp}) as candidates for entering $\mathcal{H}$ are irrelevant. Suppose that we had retained these terms. Each of them had contributed a term $w_p \tens{\lambda}^{2p}$ to the kernel in Eq.~(\ref{totalHamiltonian}). From the analysis above it is evident, however, that $w_p$ had appeared in the noise cumulants only as $w_p / w^p \sim \mu^{2-2l}$. We conclude that keeping the $w_{p\geq 2}$ had produced only corrections to scaling and that neglecting them in studying the leading behavior at the critical point is indeed justified.

\section{Renormalization group analysis}
\label{RGA}
In this section we calculate the generating function $G ( {\rm{\bf x}}, {\rm{\bf x}}^\prime , \tens{\lambda} )$ by employing field theory augmented by renormalization. For background on these methods we refer to Ref.~\cite{amit_zinn-justin}. We perform a diagrammatic perturbation calculation up to two-loop order. 

\subsection{Diagrammatic elements}
\label{diagrammaticElements}
Instead of working with ${\mathcal{H}}$ directly, we recast ${\mathcal{H}}$ into the form of a dynamic functional~\cite{janssen_dynamic,deDominicis&co,janssen_92}. This strategy is convenient because it simplifies the following calculations from the onset. Assuming $r \neq 0$ we introduce new variables by setting
\begin{eqnarray}
\label{subst}
x_\parallel = {\rm{\bf r}} \cdot {\rm{\bf x}} = r \rho t \, , \quad 
\psi = |r|^{-1/2} \, s \, , \quad
g = |r|^{1/2} \, \overline{g} \, .
\end{eqnarray}
On substituting Eq.~(\ref{subst}) into Eq.~(\ref{totalHamiltonian}) we obtain the dynamic functional
\begin{eqnarray}
\label{dynFktnal}
{\mathcal{J}} &=& \int d^{d_\perp}x_\perp  \, dt \Bigg\{ \frac{1}{2} \sum_{\tens{\lambda} \neq \tens{0}} s_{-\tens{\lambda}} \left( {\rm{\bf x}}_\perp , t \right) \bigg[ \rho \Big( \tau - \nabla^2_\perp + w \tens{\lambda}^2  
\nonumber \\
&+& \sum_{l=2}^{\infty} v_l K_l \left( \tens{\lambda} \right) \Big) + \left( \theta \left( \lambda_0 \right) - \theta \left( -\lambda_0 \right) \right) \frac{\partial}{\partial t} \bigg] s_{\tens{\lambda}} \left( {\rm{\bf x}}_\perp , t  \right)
\nonumber \\
&+& \frac{\rho \overline{g}}{6} \sum_{\tens{\lambda}, \tens{\lambda}^\prime  , \tens{\lambda} + \tens{\lambda}^\prime \neq \tens{0}} s_{-\tens{\lambda}} \left( {\rm{\bf x}}_\perp , t  \right) s_{-\tens{\lambda}^\prime} \left( {\rm{\bf x}}_\perp , t  \right) 
\nonumber \\
&\times& s_{\tens{\lambda} + \tens{\lambda}^\prime} \left( {\rm{\bf x}}_\perp , t  \right) \Bigg\} \, ,
\nonumber \\
\end{eqnarray}
where $d_\perp = d-1$. In Eq.~(\ref{dynFktnal}) we dropped a term containing a second derivative with respect to $t$ because it is irrelevant compared to the retained term containing $\partial / \partial t$. 

\begin{figure}
\epsfxsize=8.7cm
\begin{center}
\epsffile{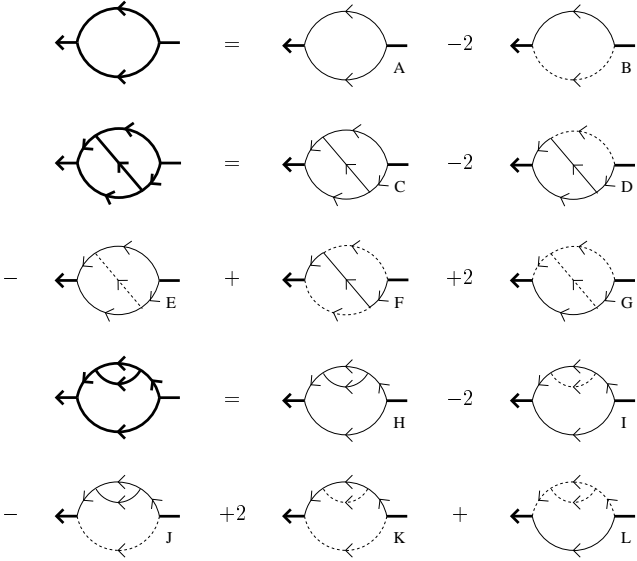}
\end{center}
\caption[]{\label{fig1}Decomposition of the primary two leg diagrams (bold) into conducting diagrams composed of conducting (light) and insulating (dashed) propagators. It is important to bear in mind that the conducting diagrams inherit their combinatorial factor from their bold diagram. For example, the diagrams A and B have to be calculated with the same combinatorial factor $\frac{1}{2}$.}
\end{figure}

From Eq.~(\ref{dynFktnal}) we gather the diagrammatic elements contributing to our renormalization group improved perturbation calculation. Dimensional analysis shows that the coupling constant $\overline{g}$ has the naive dimension $(4 - d_\perp )/2$, i.e., $d = d_\perp +1 = 5$ is the upper critical dimension. The Gaussian propagator $G ( {\rm{\bf x}}_\perp , t , \tens{\lambda} )$ is determined by the equation of motion
\begin{eqnarray}
&& \left\{ \rho \left( \tau - \nabla^2 + w \tens{\lambda}^2 \right) + \left( \theta \left( \lambda_0 \right) - \theta \left( -\lambda_0 \right) \right) \frac{\partial}{\partial t} \right\} 
\nonumber \\
&&\ \times G \left( {\rm{\bf x}}_\perp , t , \tens{\lambda} \right) = \delta \left( {\rm{\bf x}}_\perp \right) \delta \left( t \right) .
\end{eqnarray}
Note that we have not included the terms proportional to the $v_l$ due to their irrelevance. The proper treatment of these terms will be explained in Sec.~\ref{incorp}. Solving the equation of motion is straightforward. For the Fourier transform $\widetilde{G} ( {\rm{\bf p}} , t , \tens{\lambda} )$ of $G ( {\rm{\bf x}}_\perp , t , \tens{\lambda} )$ we obtain
\begin{eqnarray}
\widetilde{G} \left( {\rm{\bf p}} , t , \tens{\lambda} \right) = \widetilde{G}_+ \left( {\rm{\bf p}} , t , \tens{\lambda} \right) + \widetilde{G}_- \left( {\rm{\bf p}} , t , \tens{\lambda} \right) \, ,
\end{eqnarray}
where ${\rm{\bf p}}$ is the momentum conjugate to ${\rm{\bf x}}_\perp$ and
\begin{eqnarray}
\label{defGpm}
&&\widetilde{G}_\pm \left( {\rm{\bf p}} , t , \tens{\lambda} \right) 
\nonumber \\
&& = \, \theta \left( \pm t \right) \theta \left( \pm \lambda_0 \right) \exp \left[ \mp t \rho \left( \tau + {\rm{\bf p}}^2 + w \tens{\lambda}^2 \right) \right]
\nonumber \\
&& \times \, 
 \left( 1 - \delta_{\tens{\lambda}, \tens{0}} \right) \, .
\end{eqnarray}
For the perturbation expansion it is sufficient to keep either $\widetilde{G}_+ ( {\rm{\bf p}} , t , \tens{\lambda} )$ or $\widetilde{G}_- ( {\rm{\bf p}} , t , \tens{\lambda} )$ and hence we discard $\widetilde{G}_- ( {\rm{\bf p}} , t , \tens{\lambda} )$. The factor $( 1 - \delta_{\tens{\lambda}, \tens{0}} )$ on the right hand side of Eq.~(\ref{defGpm}) enforces the constraint $\tens{\lambda} \neq \tens{0}$. Due to this factor the principal propagator $\widetilde{G}_+ ( {\rm{\bf p}} , t , \tens{\lambda} )$ decomposes into
\begin{eqnarray}
\label{propDecomp}
\widetilde{G}_+ \left( {\rm{\bf p}} , t , \tens{\lambda} \right) = \widetilde{G}^{\text{cond}} \left( {\rm{\bf p}} , t , \tens{\lambda} \right) - \widetilde{G}^{\text{ins}} \left( {\rm{\bf p}} , t  \right)\, .
\end{eqnarray}
The first part
\begin{eqnarray}
\label{defCond}
\widetilde{G}^{\text{cond}} \left( {\rm{\bf p}} , t , \tens{\lambda} \right) &=& \theta \left(  t \right) \theta \left( \lambda_0 \right) 
\nonumber \\
&\times&
\exp \left[ - t \rho \left( \tau + {\rm{\bf p}}^2 + w \tens{\lambda}^2 \right) \right] 
\end{eqnarray}
is carrying replicated currents and hence we call it conducting.
\begin{eqnarray}
\label{defIns}
\widetilde{G}^{\text{ins}} \left( {\rm{\bf p}} , t  \right) = \theta \left(  t \right) \exp \left[ - t \rho \left(  \tau + {\rm{\bf p}}^2 \right) \right] \delta_{\tens{\lambda}, \tens{0}}
\end{eqnarray}
on the other hand is not carrying replicated currents and we refer to it as the insulating propagator. The decomposition of the principal propagator allows for a schematic decomposition of the principal diagrams into sums of conducting diagrams consisting of conducting and insulating propagators. In Fig.~\ref{fig1} we list the result of the decomposition procedure up to two-loop order.
%%%%%%%%%%%%%%%%%%%%%%%%%%%%%%%%%%%%%%%%%%%%%%%%%%%
%%%%%%%%%%%%%%%%%%%%%%%%%%%%%%%%%%%%%%%%%%%%%%%%%%%

According to our real world interpretation~\cite{stenull_janssen_epl_2000} the conducting diagrams may be viewed as being resistor networks themselves with conducting propagators corresponding to conductors and insulating propagators corresponding to open bonds. In our interpretation the times $t$ appearing in the conducting propagators correspond to resistances and the replica variables $\tens{\lambda}$ (up to a factor $-i$) to currents. Just as the physical currents are conserved in the nodes of real networks, the replica currents are conserved in each vertex and we may write for each edge $i$ of a diagram, $\tens{\lambda}_i = \tens{\lambda}_i ( \tens{\lambda} , \{ \tens{\kappa} \} )$, where $\tens{\lambda}$ is an external current and $\{ \tens{\kappa}\}$ denotes a complete set of independent loop currents.

\subsection{Diagrammatic treatment of the dangerously irrelevant couplings}
\label{incorp}
Since the coupling constants $v_l$ are irrelevant, they cannot be treated in the same fashion as the other coupling constants $\tau$ and $w$ pertaining to the bilinear part of $\mathcal{J}$. Suppose we would naively add $- t \rho \sum_l v_l K_l ( \tens{\lambda})$ to the argument of the exponential in Eq.~(\ref{defGpm}) and then use the resulting expression as the Gaussian propagator. Doing so we would ruin our perturbation expansion. This can be understood by expanding any of the diagrams in terms of $v_l$. For increasing orders of this expansion one encounters increasing orders of superficial divergence. The diagrammatic expansion can be fixed, however, by truncating the expansion in the $v_l$ at linear order. This is equivalent to treating $v_l$ by means of the insertion 
\begin{eqnarray}
\label{opdev}
\mathcal{O}^{(l)} &=& - \frac{\rho}{2} \, w^l \int d^{d_\perp} p \int dt  \, \, {\textstyle \sum_{\tens{\lambda}} } K_l \big( \tens{\lambda} \big) 
\nonumber \\
&\times& \varphi_{\tens{\lambda}} \left( {\rm{\bf p}} ,t \right) \varphi_{-\tens{\lambda}} \left( -{\rm{\bf p}} , t \right) \,  ,
\end{eqnarray}
that is associated with the coupling constant $v_l/w^l$. In Eq.~(\ref{opdev}), $\varphi_{\tens{\lambda}} \big( {\rm{\bf p}}, t \big)$ denotes the Fourier transform of $s_{\tens{\lambda}} \big( {\rm{\bf x}}_\perp ,t \big)$.

\begin{figure*}
\epsfxsize=12.5cm
\begin{center}
\epsffile{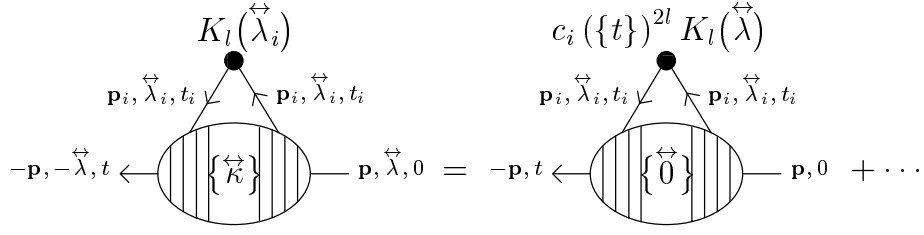}
\end{center}
\caption[]{\label{calcScheme}Calculation scheme. The hatched blobs symbolize an arbitrary number of closed conducting loops. The solid dots indicate insertions.}
\end{figure*}

Now we analyze the structure of the conducting diagrams after $\mathcal{O}^{(l)}$ has been inserted into one of their conducting propagators. This situation is sketched in Fig.~\ref{calcScheme}. Any of these diagrams has a current-dependent part of the form
\begin{eqnarray}
&& - t_i \rho w^l \sum_{\big\{ \tens{\kappa} \big\}} K_l \left( \tens{\lambda}_i \right) \exp \bigg[ - \rho w \sum_j t_j \tens{\lambda}_j^2 \bigg]
\nonumber \\
&& = \, 
- t_i \rho w^l \sum_{\big\{ \tens{\kappa} \big\}} K_l \left( \tens{\lambda}_i \right) \exp \left[ \rho w P \left( \tens{\lambda} , \left\{ \tens{\kappa} \right\} \right)  \right] \, ,
\nonumber \\
\end{eqnarray}
where $P$ denotes the electric power of the diagram. The summation is carried out by completing the squares in the exponential. The corresponding shift in the loop currents is given by the minimum of the quadratic form $P$ which is determined by a variation principle completely analogous to the one stated in Eq.~(\ref{variationPrinciple2}). Thus, completing of the squares is equivalent to solving Kirchhoff's equations for the diagram. It leads to
\begin{eqnarray}
&&- t_i \rho w^l \sum_{\big\{ \tens{\kappa} \big\}} K_l \left( \tens{\lambda}_i^{\mbox{\scriptsize ind \normalsize}} + \sum_j C_{i,j} \left( \left\{ t \right\} \right) \tens{\kappa}_j \right) 
\nonumber \\
&& \times \, 
\exp \bigg[ - \rho w R \left( \left\{ t \right\} \right) \tens{\lambda}^2 - \rho w \sum_{i,j} B_{i,j} \left( \left\{ t \right\} \right) \tens{\kappa}_i \cdot \tens{\kappa}_j \bigg] \, .
\nonumber \\
\end{eqnarray}
$\tens{\lambda}_i^{\mbox{\scriptsize ind \normalsize}} = c_i ( \{ t \} ) \tens{\lambda}$ is the current induced by the external current into edge $i$. $c_i ( \{ t \} )$ and $ C_{i,j} ( \{ t \} )$ are homogeneous functions of the times of degree zero. $B_{i,j} ( \{ t \} )$ and the total resistance of the diagram $R( \{ t \} )$ are homogeneous functions of the times of degree one. By a suitable choice of the $\tens{\kappa}_i$ the matrix constituted by the $B_{i,j}$ is rendered diagonal, i.e., $B_{i,j} \sim \delta_{i,j}$. At this stage it is convenient to switch to continuous currents and to replace the summation by an integration,
\begin{eqnarray}
\sum_{\big\{ \tens{\kappa} \big\}} \to \int \prod_{i=1}^L d \tens{\kappa}_i \ , 
\end{eqnarray}
where $L$ stands for the number of independent conducting loops. This integration is Gaussian and therefore straightforward. In the limit $D \to 0$ one obtains
\begin{eqnarray}
\label{huhu}
&& - t_i \rho w^l K_l \left( \tens{\lambda}_i^{\mbox{\scriptsize ind \normalsize}} \right) + \cdots
\nonumber \\
&&= \, - t_i c_i \left( \left\{ t \right\} \right)^{2l} \rho w^l K_l \left( \tens{\lambda} \right) + \cdots \, .
\end{eqnarray}
The terms neglected in Eq.~(\ref{huhu}) are not required in calculating the $\psi_l$. This issue is discussed in detail in Sec.~\ref{noisyRenormalization}. Diagrammatically, the calculation scheme can be condensed into Fig.~\ref{calcScheme}. Appendix~\ref{app:calculations} illustrates the calculation scheme in terms of an example.
%%%%%%%%%%%%%%%%%%%%%%%%%%%%%%%%%%%%%%%%%%%%%%%%%%%
%%%%%%%%%%%%%%%%%%%%%%%%%%%%%%%%%%%%%%%%%%%%%%%%%%%

\subsection{Renormalization and scaling}
\label{noisyRenormalization}
Now we will consider the renormalization of the field and the various parameters appearing in the dynamic functional~(\ref{dynFktnal}). Most of the techniques we are going to use, such as dimensional regularization and minimal subtraction, belong to the standard repertoire of renormalized field theory, cf.\ Ref.~\cite{amit_zinn-justin}. The renormalization of the $v_l$, however, is somewhat intricate. Hence, we will elaborate on it in this section. 

We start with those quantities that can be renormalized in a straightforward fashion. For these we use the same renormalizations as in Refs.~\cite{janssen_stenull_prerapid_2001,stenull_janssen_jsp_2001}:
\begin{subequations}
\label{renorScheme}
\begin{eqnarray}
s \to {\mathaccent"7017 s} &=& Z^{1/2} s \, ,
\\
\rho \to {\mathaccent"7017 \rho} &=& Z^{-1} Z_{\rho} \rho \, ,
\\
\tau \to {\mathaccent"7017 \tau} &=& Z_\rho^{-1} Z_{\tau} \tau \, ,
\\
w \to {\mathaccent"7017 w} &=& Z_\rho^{-1} Z_{w} w \, ,
\\
\overline{g} \to \mathaccent"7017{\overline{g}} &=& Z^{-1/2} Z^{-1}_\rho Z_u^{1/2} G_\varepsilon^{-1/2} u^{1/2} \mu^{\varepsilon /2} \, ,
\end{eqnarray}
\end{subequations}
where $\varepsilon = 4-d_\perp$ and $\mu$ is the usual inverse length scale. The factor $G_\varepsilon = (4\pi )^{-d_\perp/2}\Gamma (1 + \varepsilon /2)$, with $\Gamma$ denoting the Gamma function, is introduced for convenience. $Z$, $Z_\tau$, $Z_\rho$, and $Z_u$ are the usual DP Z-factors, which can be found to second order in $\varepsilon$ in the literature~\cite{janssen_81,janssen_2000}. $Z_w$ can be gleaned to second order in $\varepsilon$ from Refs.~\cite{janssen_stenull_prerapid_2001,stenull_janssen_jsp_2001}.

To prepare for the renormalization of the noise couplings we proceed with reviewing some general features of operator insertions. An operator ${\mathcal{O}}_i$ of a given naive dimension $\left[ {\mathcal{O}}_i \right]$ inserted one time in a vertex function generates in new primitive divergencies corresponding to operators of equal or lower naive dimension. Thus, one needs these newly generated operators as counterterms in the Hamiltonian. The operators of lower naive dimension can be isolated by additive renormalization,
\begin{eqnarray}
{\mathcal{O}}_i \to \hat{{\mathcal{O}}}_i = {\mathcal{O}}_i 
- \sum_{\left[ {\mathcal{O}}_j \right] < \left[ {\mathcal{O}}_i 
\right]} X_{i,j} {\mathcal{O}}_j \, .
\end{eqnarray}
Dimensional regularization in conjunction with minimal subtraction leads to $X_{i,j}$ containing a monomial in $\tau$ as a factor that is at least of degree one. Hence, these $X_{i,j}$ vanish at the critical point. Being interested in the leading behavior at criticality, we thus can neglect the operators of lower naive dimension in the following.

\begin{figure}
\epsfxsize=7.4cm
\begin{center}
\epsffile{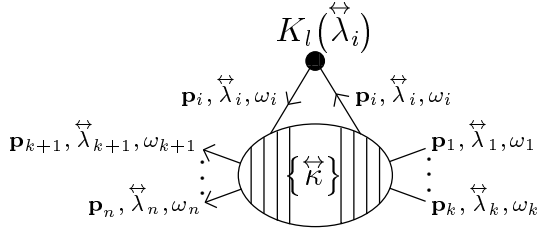}
\end{center}
\caption[]{\label{reno1}${\mathcal{O}}^{(l)}$ inserted into a conducting diagram with $n$ external legs.}
\end{figure}

Now to the particular case we are interested in, {\em viz.} the insertion of $\mathcal{O}^{(l)}$ defined in Eq.~(\ref{opdev}). Whereas the representation (\ref{opdev}) is well suited for practical calculations we find it convenient to rewrite $\mathcal{O}^{(l)}$ for the argumentation in this section as
\begin{eqnarray}
\label{oprewriten}
\mathcal{O}^{(l)} &=& - \frac{1}{2\, \rho} \, w^l \int d^{d_\perp} p \int d\omega  \, \, {\textstyle \sum_{\tens{\lambda}} } K_l \big( \tens{\lambda} \big) 
\nonumber \\
&\times& \phi_{\tens{\lambda}} \left( {\rm{\bf p}} ,\omega \right) \phi_{-\tens{\lambda}} \left( -{\rm{\bf p}} , -\omega \right) \,  ,
\end{eqnarray}
where $\phi_{\tens{\lambda}} ( {\rm{\bf p}} ,\omega )$ stands for the Fourier transform of $\varphi_{\tens{\lambda}} ( {\rm{\bf p}} ,t )$. Inserting ${\mathcal{O}}^{(l)}$ into a conducting diagram with $n$ external legs, see Fig.~\ref{reno1}, generates primitive divergencies which must be canceled by counter terms of the  
structure
\begin{eqnarray}
\label{generalOperator}
P_r \left( \tens{\lambda} \right) {\rm{\bf p}}^{2a} \omega^b \phi_{\tens{\lambda}} \left( {\rm{\bf p}} , t \right)^n \, ,
\end{eqnarray}
where
\begin{eqnarray}
P_r \left( \tens{\lambda} \right) = \prod_i K_i \left( \tens{\lambda} \right)^{r_i} \, ,
\end{eqnarray}
with $\sum_i i r_i = r$, is a homogeneous polynomial of degree $2r$. The notation that we use here and in the following is symbolic. Such a counter term depends in general on an entire set of external momenta, frequencies, and currents. $\phi_{\tens{\lambda}} ( {\rm{\bf p}} , \omega )^n$ is for example an abbreviation for $\prod_{i=1}^n \phi_{\tens{\lambda}_i} ( {\rm{\bf p}}_i , \omega_i )$. We drop constants, integrals etc.\ for notational simplicity.
%%%%%%%%%%%%%%%%%%%%%%%%%%%%%%%%%%%%%%%%%%%%%%%%%%%
%%%%%%%%%%%%%%%%%%%%%%%%%%%%%%%%%%%%%%%%%%%%%%%%%%%

\begin{figure}
\epsfxsize=6cm
\begin{center}
\epsffile{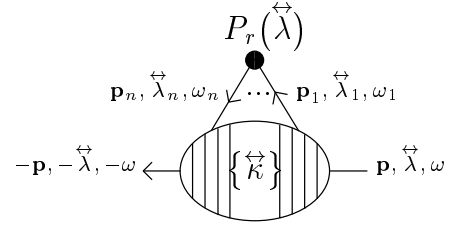}
\end{center}
\caption[]{\label{reno2}An operator of the type in Eq.~(\ref{generalOperator}) with $n \geq 3$ inserted into a conducting two-leg diagram.}
\end{figure}

The naive dimension of operators of the type (\ref{generalOperator}) is given at the upper critical dimension by $2(r+a+b+n-3)$, as straightforward power counting shows. Hence, operators having the same naive dimension as ${\mathcal{O}}^{(l)}$ satisfy
\begin{eqnarray}
\label{blabla}
l + 2 =  r + a + b + n  \, .
\end{eqnarray}
For $n=2$ one is led to $l \geq r$, i.e., the insertion of ${\mathcal{O}}^{(l)}$ generates operators containing homogeneous polynomials in the replica currents of degree equal or lower $2l$. In particular ${\mathcal{O}}^{(l)}$ generates an operator of type
\begin{eqnarray}
v_l K_l \left( \tens{\lambda} \right) \phi_{\tens{\lambda}} \left( {\rm{\bf p}} , \omega \right)^2 \, .
\end{eqnarray}
The important question now is, if the other operators generated by ${\mathcal{O}}^{(l)}$ generate ${\mathcal{O}}^{(l)}$ also. Consider $n \geq 3$. With help of Eq.~(\ref{blabla}) one obtains $l-1 \geq r \geq 1$. The second inequality is a basic feature of the summations over loop currents in the limit $D\to 0$. Bearing in mind that maximal homogeneous polynomials of degree $l-1$ in $\tens{\lambda}$ are generated, we reinsert these operators of the type in Eq.~(\ref{generalOperator}) with $n \geq 3$ into two-leg diagrams, see Fig.~\ref{reno2}. The resulting terms are of the form
\begin{eqnarray}
P_{r^\prime} \left( \tens{\lambda} \right) {\rm{\bf p}}^{2a^\prime} \omega^{b^\prime} \phi_{\tens{\lambda}} \left( {\rm{\bf p}} , \omega \right)^2 \, ,
\end{eqnarray}
with the leading contributions satisfying $r+a+b+n-3=r^\prime+a^\prime+b^\prime-1$. Thus, $r^\prime \geq r + a - a^\prime + b - b^\prime + 1$, or in other words, the homogeneous polynomials in $\tens{\lambda}$ may have a higher degree than $2l$. Nevertheless, they are of the type
\begin{eqnarray}
P_{r^\prime} \left( \tens{\lambda} \right) = K_{1} \left( \tens{\lambda} \right)^s 
\prod_{2 \leq i \leq r} K_i \left( \tens{\lambda} \right)^{r_i} \, ,
\end{eqnarray}
with $\sum_i i r_i \leq r \leq l-1$ and $\sum_i i r_i + s = r^\prime$. These polynomials have a higher symmetry than the original $K_l$. 
%%%%%%%%%%%%%%%%%%%%%%%%%%%%%%%%%%%%%%%%%%%%%%%%%%%
%%%%%%%%%%%%%%%%%%%%%%%%%%%%%%%%%%%%%%%%%%%%%%%%%%%

We have learned in the preceding paragraph that ${\mathcal{O}}^{(l)}$ generates itself and an entire family of new operators. Of course, the entire family of operators associated with ${\mathcal{O}}^{(l)}$ has to be taken into account in the renormalization procedure, leading to a renormalization in matrix form
\begin{eqnarray}
\underline{\hat{{\mathcal{O}}}}^{(l)} \to \underline{{\mathaccent"7017 {\mathcal{O}}}}^{(l)} = \left( Z_w Z_\rho^{-1} \right)^{-l} \underline{\underline{Z}}^{(l)} \underline{\hat{{\mathcal{O}}}}^{(l)} \, .
\end{eqnarray}
The vector
\begin{eqnarray}
\label{familyVector}
\underline{\hat{{\mathcal{O}}}}^{(l)} = \left( {\mathcal{O}}^{(l)}, 
\hat{{\mathcal{O}}}_2^{(l)}, \cdots \right)
\end{eqnarray}
represents the family of operators associated with ${\mathcal{O}}^{(l)}$. The factor $( Z_w Z_\rho^{-1} )^{-l}$ reflects the fact that we have incorporated a $w^l$ in the definition of ${\mathcal{O}}^{(l)}$. 

The important conclusion from our reasoning above is that the operators generated by ${\mathcal{O}}^{(l)}$ do not in turn generate ${\mathcal{O}}^{(l)}$. For operators with this outstanding feature we have introduced the notion of master operators~\cite{stenull_janssen_epl_2000}. Master operators are associated with renormalization matrices
\begin{eqnarray}
\underline{\underline{Z}}^{(l)} = \underline{\underline{1}} + O \left( u \right) \, ,
\end{eqnarray}
where $\underline{\underline{1}}$ stands for the unit matrix, of a particularly simple structure,
\begin{eqnarray}
\underline{\underline{Z}}^{(l)} = 
\left(
\begin{array}{cccc}
Z^{(l)} & \Diamond & \cdots & \Diamond \\
0       & \Diamond & \cdots & \Diamond \\
\vdots  & \vdots & \ddots & \vdots \\
0       & \Diamond & \cdots & \Diamond
\end{array}
\right) \, .
\end{eqnarray}
The $\Diamond$ symbolizes arbitrary elements. We will see in Sec.~\ref{noisyScaling}, as a consequence of the simple structure of $\underline{\underline{Z}}^{(l)}$, that the operators induced by ${\mathcal{O}}^{(l)}$ can be neglected in calculating the scaling index of their master. Owing to their subordinate role, we refer to these operators as servants.

\subsection{Scaling}
\label{noisyScaling}
In this section we set up a Gell-Mann--Low renormalization group equation (RGE). Its solution will provide us with the scaling behavior of the order parameter correlation functions and finally with the scaling behavior of the $C_R^{(l)}$.

The bare (unrenormalized) theory has to be independent of the length scale $\mu^{-1}$ introduced by renormalization. Thus, the bare connected $N$ point correlation functions satisfy the identity  
\begin{eqnarray}
\label{independence}
\mu \frac{\partial}{\partial \mu} {\mathaccent"7017 G}_N \left( \left\{ {\rm{\bf x}}_\perp , {\mathaccent"7017 \rho} t , {\mathaccent"7017 w} \tens{\lambda}^2 \right\} ; {\mathaccent"7017 \tau}, \mathaccent"7017{\overline{g}} \right)_{\underline{\hat{{\mathcal{O}}}}^{(l)}} = 0 \, .
\end{eqnarray}
The subscript $\underline{\hat{{\mathcal{O}}}}^{(l)}$ indicates that the corresponding operator has been inserted. Eq.~(\ref{independence}) translates via the Wilson functions defined by
\begin{subequations}
\begin{eqnarray}
\label{wilson}
\gamma_{...} \left( u \right) &=& \mu \frac{\partial }{\partial \mu} \ln Z_{...}  \bigg|_0 \, ,
\\
\beta \left( u \right) &=& \mu \frac{\partial u}{\partial \mu} \bigg|_0 = \left( - \varepsilon + \gamma + 2 \gamma_\rho - \gamma_u \right) u \, ,
\nonumber \\
\\ 
\kappa \left( u \right) &=& \mu \frac{\partial
\ln \tau}{\partial \mu}  \bigg|_0 = \gamma_\rho - \gamma_\tau \, ,
 \\
\zeta_w \left( u \right) &=& \mu \frac{\partial \ln w}{\partial \mu}  \bigg|_0 = \gamma_\rho - \gamma_w \, ,
\\
\zeta_\rho \left( u \right) &=& \mu \frac{\partial \ln \rho}{\partial \mu}  \bigg|_0 = \gamma - \gamma_\rho \, ,
\\
\underline{\underline{\gamma}}^{(l)} \left( u \right) &=& - \mu \frac{\partial }{\partial \mu} \ln \underline{\underline{Z}}^{(l)}  \bigg|_0 \, ,
\end{eqnarray}
\end{subequations}
(the $|_0$ indicates that bare quantities are kept fix while taking the derivatives) into the RGE
\begin{eqnarray}
&&\left\{ \left[ \mathcal{D}_\mu + \frac{N}{2} \gamma \right] \underline{\underline{1}} + \underline{\underline{\gamma}}^{(l)} - l \zeta_w \, \underline{\underline{1}} \right\}
\nonumber \\
&&\, \times
G_N \left( \left\{ {\rm{\bf x}}_\perp , \rho t  ,w \tens{\lambda}^2 \right\} ; \tau, u, \mu \right)_{\underline{\hat{{\mathcal{O}}}}^{(l)}} = 0 \, .
\nonumber \\
\end{eqnarray}
Here, $\mathcal{D}_\mu$ is a shorthand for
\begin{eqnarray}
\mathcal{D}_\mu = \mu \frac{\partial }{\partial \mu} + \beta \frac{\partial }{\partial u} + \tau \kappa \frac{\partial }{\partial \tau} + w \zeta_w \frac{\partial }{\partial w} + \rho \zeta_\rho \frac{\partial }{\partial \rho}  \, .
\nonumber \\
\end{eqnarray}

To solve the RGE we employ the method of characteristics. Considering the ingredients of the RGE as being functions of a single flow parameter $\ell$ we write
\begin{subequations}
\begin{eqnarray}
\ell \frac{\partial \bar{\mu}}{\partial 	\ell} &=& \bar{\mu} \, , \quad \bar{\mu}(1)=\mu \ ,
\\
\label{charBeta}
\ell \frac{\partial \bar{u}}{\partial \ell} &=& \beta \left( \bar{u}(\ell ) \right) \, , \quad \bar{u}(1)=u \, ,
\\
\ell \frac{\partial}{\partial \ell} \ln \bar{\rho} &=& \zeta_\rho \left( \bar{u}(\ell ) \right) \, , \quad \bar{\rho}(1)=\rho \, ,
\\
\ell \frac{\partial}{\partial \ell} \ln \bar{\tau} &=& \kappa \left( \bar{u}(\ell ) \right) \, , \quad \bar{\tau}(1)=\tau \, ,
\\
\ell \frac{\partial}{\partial \ell} \ln \bar{w} &=& \zeta_w \left( \bar{u}(\ell ) \right) \, , \quad \bar{w}(1)=w \, ,
\\
\ell \frac{\partial}{\partial \ell} \ln \bar{Z} &=& \gamma \left( \bar{u}(\ell ) \right) \, , \quad \bar{Z}(1)=1 \, ,
\\
\ell \frac{\partial}{\partial \ell} \ln \bar{\underline{\underline{Z}}}^{(l)} &=&- \underline{\underline{\gamma}}^{(l)}  \left( \bar{u}(\ell ) \right) \, , \quad \bar{\underline{\underline{Z}}}(1)= \underline{\underline{1}} \, .
\end{eqnarray}
\end{subequations}
These characteristics describe how the parameters transform if we change the momentum scale $\mu $ according to $\mu \to \bar{\mu}(\ell )=\ell \mu $. Being interested in the infrared (IR) behavior of the theory, we study the limit $\ell \to 0$. According to Eq.~(\ref{charBeta}) we expect that in this IR limit the coupling constant $\bar{u}(\ell )$ flows to a stable fixed point $u^\ast$ satisfying $\beta (u^\ast )=0$. At this fixed point the RGE simplifies to 
\begin{eqnarray}
\label{fixedPointRGE}
&&\left\{ \left[ \mathcal{D}_\mu^\ast + \frac{N}{2} \gamma^\ast \right] \underline{\underline{1}} + \underline{\underline{\gamma}}^{(l)\ast} - l \zeta_w^\ast \, \underline{\underline{1}} \right\}
\nonumber \\
&&\, \times
G_N \left( \left\{ {\rm{\bf x}}_\perp , \rho t  ,w \tens{\lambda}^2 \right\} ; \tau, u^\ast , \mu \right)_{\underline{\hat{{\mathcal{O}}}}^{(l)}} = 0 \, ,
\nonumber \\
\end{eqnarray}
where $\gamma^\ast$ is an abbreviation for $\gamma (u^\ast)$, $\underline{\underline{\gamma}}^{(l)\ast}$ for $\underline{\underline{\gamma}}^{(l)} (u^\ast)$, and so on. To proceed towards a solution of the RGE it is important to realize that the matrix $\underline{\underline{\gamma}}^{(l)\ast}$ inherits the simple structure of $\underline{\underline{Z}}^{(l)}$, {\em viz.}
\begin{eqnarray}
\underline{\underline{\gamma}}^{(l)\ast} = 
\left(
\begin{array}{cccc}
\gamma^{(l)\ast} & \Diamond & \cdots & \Diamond \\
0       & \Diamond & \cdots & \Diamond \\
\vdots  & \vdots & \ddots & \vdots \\
0       & \Diamond & \cdots & \Diamond
\end{array}
\right) \, .
\end{eqnarray}
By virtue of this structure, $\left| 1 \right\rangle = \left( 1, 0, \cdots, 0 \right)^T$ is a right eigenvector with eigenvalue $\gamma^{(l)\ast}$. We denote the remaining right eigenvectors with eigenvalues $\gamma_{k\geq 2}^{(l)\ast}$ by $\left| k \right\rangle$. The left eigenvectors are $\left\langle 1 \right| = \left( 1, \Diamond, \cdots , \Diamond \right)$ and $\left\langle k \right| = \left( 0, \Diamond, \cdots , \Diamond \right)$. Employing spectral decomposition we recast $\underline{\underline{\gamma}}^{(l)\ast}$ in terms of its eigenvalues and eigenvectors as
\begin{eqnarray}
\label{spectralDeco}
\underline{\underline{\gamma}}^{(l)\ast} = \left| 1 \right\rangle \gamma^{(l)\ast} \left\langle 1 \right| + \sum_{k \geq 2} \left| k \right\rangle \gamma^{(l)\ast}_k \left\langle k \right| \, .
\end{eqnarray}
Now we substitute the decomposition (\ref{spectralDeco}) into the RGE~(\ref{fixedPointRGE}). Multiplying the resulting equation from the left hand side with $\langle 1|$ leads to
\begin{eqnarray}
\label{SimpleFixedPointRGE}
&&\left[ \mathcal{D}_\mu^\ast + \frac{N}{2} \gamma^\ast + \gamma^{(l)\ast} - l \zeta_w^\ast \right]
\nonumber \\
&&\, \times
G_N \left( \left\{ {\rm{\bf x}}_\perp , \rho t  ,w \tens{\lambda}^2 \right\} ; \tau, u^\ast , \mu \right)_{\mathcal{A}^{(l)}} = 0 \, ,
\nonumber \\
\end{eqnarray}
Here, $\mathcal{A}^{(l)}$ is an abbreviation for
\begin{eqnarray}
{\mathcal{A}}^{(l)} = \left\langle 1 \right| \underline{\hat{\mathcal{O}}}^{(l)} &=& \hat{\mathcal{O}}^{(l)} + \sum_{k \geq 2} \Diamond \, \hat{\mathcal{O}}^{(l)}_k \, .
\end{eqnarray}
Note that
\begin{eqnarray}
\left\langle k \right| \underline{\hat{\mathcal{O}}}^{(l)} &=&  \sum_{m \geq 
2} \Diamond \, \hat{\mathcal{O}}^{(l)}_m \, ,
\end{eqnarray}
i.e., the RGE~(\ref{SimpleFixedPointRGE}) contains all the information on the ${\mathcal{O}}^{(l)}$. In the from~(\ref{SimpleFixedPointRGE}) the RGE is readily solved. With help of the characteristics we obtain
\begin{eqnarray}
\label{SolOfRgg}
&&G_N \left( \left\{ {\rm{\bf x}}_\perp , \rho t ,w \tens{\lambda}^2 \right\} ; \tau, u, \mu \right)_{\mathcal{A}^{(l)}} 
\nonumber \\
&&\, = \ell^{\gamma^\ast N/2 + \gamma^{(l)\ast} - l\zeta_w^\ast} 
\nonumber \\
&&\, \times
G_N \left( \left\{ \ell{\rm{\bf x}}_\perp ,\ell^{\zeta_\rho^\ast}\rho t , \ell^{\zeta_w^\ast}w \tens{\lambda}^2 \right\} ; \ell^{\kappa^\ast}\tau , u^\ast, \ell \mu \right)_{\mathcal{A}^{(l)}} \, .
\nonumber \\ 
\end{eqnarray}
To account for the naive dimensions of the various quantities, Eq.~(\ref{SolOfRgg}) needs to be supplemented by a dimensional analysis. Simple power counting reveals that
\begin{eqnarray}
\label{dimAna}
&&G_N \left( \left\{ {\rm{\bf x}}_\perp ,\rho t ,w \tens{\lambda}^2 \right\} ; \tau, u, \mu \right)_{\mathcal{A}^{(l)}} 
\nonumber \\
&&\, = 
\mu^{d_\perp N/2 + 2l -2} 
\nonumber \\
&&\, \times
G_N \left( \left\{ \mu {\rm{\bf x}}_\perp , \mu^2 \rho t , \mu^{-2}w \tens{\lambda}^2 \right\} ; \mu^{-2}\tau , u, 1 \right)_{\mathcal{A}^{(l)}} \, .
\nonumber \\ 
\end{eqnarray}
Equation~(\ref{SolOfRgg}) in conjunction with Eq.~(\ref{dimAna}) now gives
\begin{eqnarray}
\label{scaling}
&&G_N \left( \left\{ {\rm{\bf x}}_\perp , \rho t ,w \tens{\lambda}^2 \right\} ; \tau, u, \mu \right)_{\mathcal{A}^{(l)}}
\nonumber \\
&&\, = 
\ell^{(d_\perp +\eta)N/2 - \psi_l/\nu_\perp + l\phi/\nu_\perp} 
\nonumber \\
&&\, \times
G_N \left( \left\{ \ell{\rm{\bf x}}_\perp , \ell^z \rho t , \ell^{-\phi/\nu_\perp}w \tens{\lambda}^2 \right\} ; 
\ell^{-1/\nu_\perp}\tau , u^\ast, \mu \right)_{\mathcal{A}^{(l)}} \, .
\nonumber \\
\end{eqnarray}
Equation~(\ref{scaling}) features the well known critical exponents for DP that have been calculated previously to second order in $\varepsilon$~\cite{janssen_81,janssen_2000}:
\begin{subequations}
\begin{eqnarray}
\eta &=& \gamma^\ast = - \frac{\varepsilon}{6} \left\{ 1 + \left[ \frac{25}{288} + \frac{161}{144}\ln \left( \frac{4}{3} \right) \right] \varepsilon \right\} \, ,
\nonumber \\
\\
z &=& 2 + \zeta_\rho^\ast = 2 - \frac{\varepsilon}{12} \left\{ 1 + \left[ \frac{67}{288} + \frac{59}{144}\ln \left( \frac{4}{3} \right) \right] \varepsilon \right\} \, ,
\nonumber \\
\\
\nu_\perp &=& \frac{1}{2-\kappa^\ast } = \frac{1}{2} + \frac{\varepsilon}{16} \left\{ 1 + \left[ \frac{107}{288} - \frac{17}{144}\ln \left( \frac{4}{3} \right) \right] \varepsilon \right\} \, .
\nonumber \\
\end{eqnarray}
\end{subequations}
$\phi = \nu_\perp \left( 2 - \zeta_w^\ast \right)$ is the resistance exponent for DP that we derived recently~\cite{janssen_stenull_prerapid_2001,stenull_janssen_jsp_2001} to second order in $\varepsilon$:
\begin{eqnarray}
\phi = 1 + \frac{\varepsilon}{24} \left\{ 1 + \left[ \frac{151}{288} - \frac{157}{144}\ln \left( \frac{4}{3} \right) \right] \varepsilon \right\} \, .
\end{eqnarray}
$\psi_l$ is defined by $\psi_l = \nu ( 2 - \gamma^{(l)\ast} )$. The $\varepsilon$ expansion result of $\psi_l$ is given below.

From here only a few more steps are required to reveal the scaling behavior of the $C^{(l)}_R$. Recall that our strategy is to derive the $C^{(l)}_R$ from their generating function $G ( {\bf x}, {\bf x}^\prime ; \tens{\lambda} )$. By now, we know of the scaling behavior of a central ingredient to the generating function, {\em viz.} we know that the two-point correlation function with insertion scales at criticality as
\begin{eqnarray}
\label{twoPointScaling}
&&G_2 \left( |{\rm{\bf x}}_\perp-{\rm{\bf x}}_\perp^\prime |, t-t^\prime , w \tens{\lambda}^2 \right)_{\mathcal{A}^{(l)}} 
\nonumber \\
&& \, = 
\ell^{d_\perp + \eta - \psi_l/\nu_\perp + l\phi/\nu_\perp} 
\nonumber \\
&& \, \times
G_2 \left( \ell |{\rm{\bf x}}_\perp-{\rm{\bf x}}_\perp^\prime|,\ell^z \left( t - t^\prime \right) , \ell^{-\phi/\nu_\perp} w \tens{\lambda}^2 \right)_{\mathcal{A}^{(l)}} \, ,
\nonumber \\
\end{eqnarray}
where we dropped several arguments for notational simplicity. In the following we set ${\rm{\bf x}}_\perp^\prime = {\rm{\bf 0}}$ and $t^\prime = 0$, again for the sake of simplicity. A further ingredient to the generating function is the two-point correlation function without insertion. Its scaling behavior can be inferred from a renormalization group treatment similar to that above. This analysis is comparatively straightforward (cf. Ref.~\cite{stenull_janssen_jsp_2001}) and gives
\begin{eqnarray}
\label{scaleRel}
&&G_2 \left( |{\rm{\bf x}}_\perp |, t , w \tens{\lambda}^2 \right) 
\nonumber \\
&& \, = 
\ell^{d_\perp +\eta} G_2 \left( \ell |{\rm{\bf x}}_\perp|,\ell^z  t  , \ell^{-\phi/\nu_\perp} w \tens{\lambda}^2 \right) \, .
\nonumber \\
\end{eqnarray}
Now we put Eqs.~(\ref{twoPointScaling}) and (\ref{scaleRel}) together. Recalling that our master operator ${\mathcal{O}}^{(l)}$ is associated with a coupling constant $v_l/w^l$ we write the generating function as
\begin{eqnarray}
\label{erzeuger}
&&G \left( |{\rm{\bf x}}_\perp |, t ,  \tens{\lambda} \right) =  \ell^{d_\perp +\eta}  
\bigg\{ G_2 \left( \ell |{\rm{\bf x}}_\perp|,\ell^z t  , \ell^{-\phi/\nu_\perp} w \tens{\lambda}^2 \right) 
\nonumber \\
&& \, + \sum_{l=2}^\infty \frac{v_l}{w^l} \, \ell^{- \psi_l/\nu_\perp + l\phi/\nu_\perp}
\nonumber \\
&& \, \times
G_2 \left( \ell |{\rm{\bf x}}_\perp|,\ell^z  t  , \ell^{-\phi/\nu_\perp} w \tens{\lambda}^2 \right)_{\mathcal{A}^{(l)}} \bigg\} \, .
\end{eqnarray}
We have the freedom to choose the flow parameter as we are pleased. The choice $\ell = |{\rm{\bf x}}_\perp|^{-1}$ and a Taylor expansion of the right hand side of Eq.~(\ref{erzeuger}) lead to 
\begin{eqnarray}
&&G \left( |{\rm{\bf x}}_\perp|, t ,  \tens{\lambda} \right) = \left| {\rm{\bf x}}_\perp \right|^{1-d-\eta} f \left(  \frac{t}{\left| {\rm{\bf x}}_\perp \right|^z} \right) 
\nonumber \\
&& \times \, 
\bigg\{ 1 + w \tens{\lambda}^2 \left| {\rm{\bf x}}_\perp \right|^{\phi /\nu_\perp} f_{w} \left(  \frac{t}{\left| {\rm{\bf x}}_\perp \right|^z} \right) \nonumber \\
&& + \sum_{l=2}^\infty v_l K_l \left( \tens{\lambda} \right) \left| {\rm{\bf x}}_\perp \right|^{\psi_l /\nu_\perp} f_{v_l} \left(  \frac{t}{\left| {\rm{\bf x}}_\perp \right|^z} \right) + \cdots \bigg\} \, ,
\nonumber \\
\end{eqnarray}
where the $f$s are scaling functions that vanish for vanishing argument. Instead of choosing $\ell = |{\rm{\bf x}}_\perp|^{-1}$ we can likewise choose $\ell=t^{-1/z}$. This leads then upon Taylor expansion to
\begin{eqnarray}
&&G \left( |{\rm{\bf x}}_\perp|, t ,  \tens{\lambda} \right) = t^{(1-d-\eta)/z} \, h \left(  \frac{\left| {\rm{\bf x}}_\perp \right|^z}{t} \right) 
\nonumber \\
&& \times \, 
\bigg\{ 1 + w \tens{\lambda}^2 \, t^{\phi /\nu_\parallel} \, h_{w} \left(  \frac{\left| {\rm{\bf x}}_\perp \right|^z}{t} \right) 
\nonumber \\
&& + \sum_{l=2}^\infty v_l K_l \left( \tens{\lambda} \right) \, t^{\psi_l /\nu_\parallel} \, h_{v_l} \left(  \frac{\left| {\rm{\bf x}}_\perp \right|^z}{t} \right) + \cdots \bigg\} \, ,
\nonumber \\
\end{eqnarray}
with the $h$s being scaling functions that tend to constants for vanishing arguments. $\nu_\parallel$ is defined by $\nu_\parallel = \nu_\perp z$.

Possessing of the generating function we solely need to take the appropriate derivatives to extract the scaling behavior of the $C_R^{(l)}$ that is by virtue of Eq.~(\ref{finalCumulant}), up to unimportant constants, identical to that of the $M_I^{(l)}$. With help of Eq.~(\ref{exploitGenFkt}) we deduce that
\begin{eqnarray} 
M_I^{(l)} \sim t^{\psi_l /\nu_\parallel}
\end{eqnarray}
if measured along the preferred direction. For measurements in other directions it is appropriate to choose a length scale $L$ and to express the longitudinal and the transverse coordinates in terms of $L$: $\left| {\rm{\bf x}}_\perp \right| \sim L$ and $x_\parallel \sim L^z \sim T$. With this choice the scaling functions $f$ reduce to constants and we obtain 
\begin{eqnarray} 
M_I^{(l)} \sim L^{\psi_l /\nu_\perp} \sim T^{\psi_l /\nu_\parallel} \, .
\end{eqnarray}

\begin{figure}
\epsfxsize=8cm
\begin{center}
\epsffile{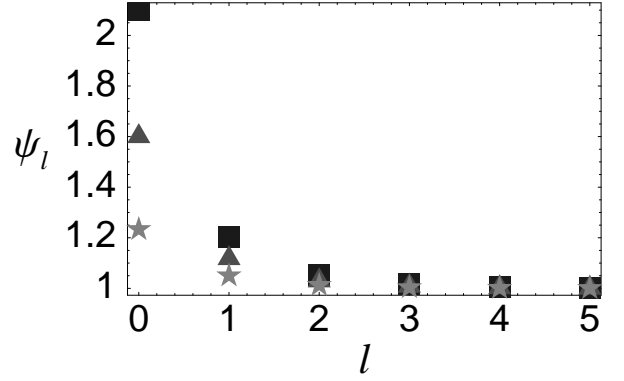}
\end{center}
\caption[]{\label{exponents}Dependence of $\psi_l$ on $l$ for $d_\perp =3$ (stars),  $d_\perp =2$ (triangles), and  $d_\perp =1$ (squares).}
\end{figure}

We still owe our result for the multifractal exponents. Since we only need to compute a single element of each of the renormalization matrices $\underline{\underline{Z}}^{(l)}$, {\em viz.} $Z^{(l)}$, we manage to calculate the $\psi_l$ to two-loop order. In $\varepsilon$-expansion, our result reads
\begin{eqnarray}
\label{monsterExponent}
\psi_l = 1 + \frac{\varepsilon}{3 \cdot 2^{2l+1}} + \varepsilon^2 \left[ a(l) - b(l) \, \ln \left( \frac{4}{3} \right) \right] + O \left( \varepsilon^3 \right) \, .
\nonumber \\
\end{eqnarray}
The $a(l)$ and $b(l)$ are $l$-dependent coefficients taking on the values listed in Table~\ref{tab:coeffs}.
%%%%table%%%%%
\begin{table*}
\caption{The coefficients $a(l)$ and $b(l)$ appearing in Eq.~\ref{monsterExponent}.}
\label{tab:coeffs}
\begin{tabular}{c||c|c|c|c|c|c|c|c}
\hline \hline
$\quad l \quad $ & $0$ & $1$ & $2$ & $3$ & $4$ & $5$ & $6$ & $\geq 7$\\ \hline
$ a(l) $ & $\frac{85}{1728}$ & $\frac{151}{6912}$ & $\frac{68387}{4976640}$ & $\frac{3307921}{334430208}$ & $\frac{4661703289}{619173642240}$ & $\frac{8258257317517}{1373079469031424} $ & $\frac{24071498466367}{4808089723207680}$ & $0.005 > a > 0$\\
\hline
$ b(l) $ & $-\frac{53}{864}$ & $\frac{157}{3456}$ & $\frac{1091}{27648}$ & $\frac{13589}{442368}$ & $\frac{173149}{7077888}$ & $\frac{2281853}{113246208}$ & $\frac{30950909}{1811939328}$ & $0.015 > b > 0$\\
\hline \hline
\end{tabular}
\end{table*}
%%%%%%%%%%%%%
$\psi_0$ and $\psi_1$ stem from extending the sum over $l$ in the Hamiltonian~(\ref{totalHamiltonian}) so that it comprises $l=0$ and $l=1$. Figure~\ref{exponents} depicts the dependence of $\psi_l$ on $l$ for $\varepsilon = 1, 2, 3$ corresponding to $d_\perp = 3, 2, 1$.
%%%%%%%%%%%%%%%%%%%%%%%%%%%%%%%%%%%%%%%%%%%%%%%%%%%
%%%%%%%%%%%%%%%%%%%%%%%%%%%%%%%%%%%%%%%%%%%%%%%%%%%

We point out that our result~(\ref{monsterExponent}) fulfills several consistency checks. $\psi_0$ is related to the fractal dimension $D_B$ of DP clusters via $D_B = 1 + \psi_0 /\nu_\perp -z$ (cf.~\cite{janssen_stenull_prerapid_2001,stenull_janssen_pre_2001_nonlinear}). Equation~(\ref{monsterExponent}) is in agreement with the $\varepsilon$-expansions of $\nu_\perp$, $z$~\cite{janssen_81,janssen_2000}, and $D_B$ (see Refs.~\cite{janssen_stenull_prerapid_2001,stenull_janssen_pre_2001_nonlinear}) to second order in $\varepsilon$. $\psi_1$ is in conformity with our result for the resistance exponent $\phi$ given in Refs.~\cite{janssen_stenull_prerapid_2001,stenull_janssen_jsp_2001}. This has to be the case because $C^{(1)}_R = M^{(1)}_R$. Multifractal exponents like the $\psi_l$ have the general feature that they are convex monotonically decreasing if being understood as a function of the index $l$~\cite{duplantier_ludwig_91}. Our result possesses of this feature. Moreover, it tends to unity for large $l$ as one expects from the relation of $\psi_\infty$ to the fractal dimension of the singly connected (red) bonds, $d_{\text{red}} = 1 + \psi_\infty /\nu_\perp -z$, see Refs.~\cite{arcangelis_etal_85,coniglio_81_82,stenull_janssen_pre_2001_nonlinear}.

\section{Concluding remarks}
\label{conclusions}
In summary, we derived a field theoretic Hamiltonian for RDN that captures the multifractality of the current distribution in these networks. To characterize the current distribution, we determined the scaling behavior of its moments. Each moment is governed by an independent critical exponent, i.e., these exponents are not related to each other in a linear or affine fashion, as commonly occurs in critical phenomena under the name of gap scaling. We determined the family of multifractal moments to two-loop order. 

Our approach thrived on two cornerstones, {\em viz.} our real-world interpretation of Feynman diagrams and our concept of master operators. The real-world interpretation remedies the apparent complexity of the field theory. It makes the field theory more intuitive and provides practical guidance for the diagrammatic calculations. Being interested in some quantity in real networks one basically just has to determine its counterpart in the Feynman diagrams. In the present case we determined the multifractal moments of the diagrams to study the multifractal moments in physical RDN. Without the concept of master operators the renormalization group analysis presented in this paper is hardly feasible. Since the multifractal moments correspond in the field theoretic formulation to dangerously irrelevant operators $\mathcal{O}^{(l)}$ they generate under renormalization a myriad of other irrelevant operators. All these must be taken into account in the renormalization group. Thus, on has in principle to compute and diagonalize renormalization matrices that are giants for large $l$. Already handling the full renormalization matrix associated with $l=2$ to one-loop order is tedious. The effort is comparable to that of determining corrections to scaling associated with a $(\vec{\lambda}^2)^2$ term in the field theory of RRN~\cite{janssen_stenull_corrections}. Due to the master operator property of the $\mathcal{O}^{(l)}$, however, it is sufficient for our purposes to calculate a single element of the renormalization matrix pertaining to each $\mathcal{O}^{(l)}$, and we can work to two-loop order with reasonable effort. To date, the concept of master operators has proved to be powerful in studying multifractality in RRN and RDN. We expect, though, that it has many more applications. It might be the case, that any multifractal quantity can be associated in the field theoretic framework with master operators. This is a speculation, but it is not implausible at all. For example, preliminary studies of the random field Ising model indicate the applicability of the master operator concept.

To our knowledge, the issue of multifractality in DP has not been studied hitherto. In particular, we do not know of any other work, theoretical, simulational or experimental, that provides results suitable for comparison to ours. For the RRN, in contrast, the multifractal exponents of the current distribution have been determined by Monte Carlo simulations~\cite{rammal_etal_prl_85,batrouni&co_96}. It would be very interesting to have corresponding numerical estimates for the RDN. According work is in progress~\cite{hinrichsen_stenull_janssen_01} and will be reported in the near future.

\begin{acknowledgments}
This work has been supported by the Sonderforschungsbereich 237 ``Unordnung und gro{\ss}e Fluktuationen'' of the Deutsche Forschungsgemeinschaft. Also, O. S. acknowledges support by the Emmy-Noether-Programm of the Deutsche Forschungsgemeinschaft.
\end{acknowledgments}

%appendices
\appendix
\section{Calculation scheme for diagrams with insertion}
\label{app:calculations}
In this Appendix we illustrate our calculation scheme outlined in Sec.~\ref{incorp} in terms of an example. For the sake of simplicity we consider the simplest conducting diagram comprising a closed loop of conducting propagators, namely diagram A introduced in Fig.~\ref{fig1}. With $\mathcal{O}^{(2)}$ inserted successively in both conducting propagators the mathematical expression for that diagram reads
\begin{eqnarray}
\label{A1}
&&\mbox{A}_{\mathcal{O}^{(2)}} = \frac{\rho^2 g^2}{2} \int_0^\infty dt \int_{\brm{k}} \sum_{\tens{\kappa}} \exp \left( - i \omega t \right)
\nonumber \\
&& \times \, \exp \left\{ -\rho t \left[ 2\tau + \brm{k}^2 + \left( \brm{k} - \brm{p} \right)^2 + w \tens{\kappa}^2 + w \big( \tens{\kappa} - \tens{\lambda} \big)^2 \right] \right\} 
\nonumber \\
&& \times \,
(-) \rho w^2 t \Big\{ K_l \left( \tens{\kappa} \right) + K_l \left( \tens{\lambda} - \tens{\kappa} \right) \Big\}
\, ,
\end{eqnarray}
where $\int_{\brm{k}}$ is an abbreviation for $(2\pi)^{-d_\perp} \int d^{d_\perp}k$. Note that $P (\tens{\lambda}, \tens{\kappa}) = -t \tens{\kappa}^2 - t ( \tens{\kappa} - \tens{\lambda} )^2$ corresponds to the electric power of diagram A. For practical purposes we switch to continuous loop currents. This step is justified at this stage because the constraint $\tens{\lambda} \neq \tens{0}$ is safely implemented via the decomposition of the corresponding bold diagram into its conducting diagrams. We obtain
\begin{eqnarray}
\label{A2}
&&\mbox{A}_{\mathcal{O}^{(2)}} = - \frac{\rho g^2}{2} \int_0^\infty dt \, t\exp \left( - i \frac{\omega}{\rho} t \right) 
\nonumber \\
&& \times \,
\int_{\brm{k}} \exp \left[ - t \left[ 2\tau + \brm{k}^2 + \left( \brm{k} - \brm{p} \right)^2 \right] \right]
\nonumber \\
&& \times \,
\int_{-\infty}^\infty d \tens{\kappa} \,  \exp \left[ w P \left( \tens{\lambda}, \tens{\kappa} \right) \right] 
\nonumber \\
&& \times \,
 w^2  \bigg\{ K_l \left( \tens{\kappa} \right) + K_l \left( \tens{\lambda} - \tens{\kappa} \right) \bigg\}
\, ,
\end{eqnarray}
where we have also modified the integration variable $t$. The integration over the loop current can be simplified by completing the squares in the exponential. We look for the minimum of the quadratic form $P (\tens{\lambda}, \tens{\kappa})$. The minimum is determined by a variation principle completely analogous to the one stated in Eq.~(\ref{variationPrinciple2}). Thus, completing the squares is equivalent to solving Kirchhoff's equations for the diagram. After carrying out the straightforward momentum integration we then find
\begin{eqnarray}
\label{A3}
&&\mbox{A}_{\mathcal{O}^{(2)}} = - \frac{\rho g^2}{2} \frac{1}{(4\pi)^{d_\perp /2}} \int_0^\infty dt \, t \, (2t)^{-d_\perp /2} 
\nonumber \\
&& \times \,
\exp \left[ - t \bigg( \frac{i\omega}{\rho}   + 2 \tau + \frac{1}{2} \brm{p}^2  \bigg) \right] 
\nonumber \\
&& \times \,
\exp \left[ - R(t) w \tens{\lambda}^2 \right] \int_{-\infty}^\infty d \tens{\kappa} \,  \exp \left[ - t w  \tens{\kappa}^2  \right] 
\nonumber \\
&& \times \,
 w^2  \Bigg\{ K_l \left( \tens{\kappa} + \frac{1}{2} \tens{\lambda}\right) + K_l \left( \tens{\kappa} - \frac{1}{2} \tens{\lambda} \right) \Bigg\}
\, .
\end{eqnarray}
$R(t) = t/2$ is the total resistance of diagram A. $\tens{\lambda}/2$, and, respectively $-\tens{\lambda}/2$, are the currents induced in the conducting propagators by the external current. Upon integrating out the loop current we get in the replica limit $D\to 0$
\begin{eqnarray}
\label{A4}
&&\mbox{A}_{\mathcal{O}^{(2)}} = - \frac{\rho g^2}{8} \frac{1}{(4\pi)^{d_\perp /2}} \int_0^\infty dt \, t^{1-d_\perp /2} 
\nonumber \\
&& \times \,
\exp \left[ - t \bigg( \frac{i \omega}{2\rho}   +  \tau + \frac{1}{4} \brm{p}^2 + \frac{w}{4} \tens{\lambda}^2 \bigg) \right] 
\nonumber \\
&& \times \,
\Bigg\{ \frac{1}{8} w^2  K_l \left( \tens{\lambda} \right) + \frac{w}{t} \tens{\lambda}^2 \Bigg\}
\, .
\end{eqnarray}
Next we expand the exponential function. Then we carry out the remaining integration. Upon discarding convergent terms that are not required for renormalization purposes we obtain in $\varepsilon$-expansion
\begin{eqnarray}
\label{A5}
&&\mbox{A}_{\mathcal{O}^{(2)}} = -\rho g^2 \frac{G_\varepsilon}{32\, \varepsilon} \, \tau^{-\varepsilon /2} 
\bigg\{  w^2  K_l \left( \tens{\lambda} \right) 
\nonumber \\
&&
- \, w \tens{\lambda}^2 \bigg[ 4\frac{i \omega}{\rho} + 8 \tau + 2 \left( \brm{p}^2 + w \tens{\lambda}^2 \right) \bigg] \bigg\} \, .
\end{eqnarray}
The example considered here highlights two points. First, not only primitive divergencies proportional to $K_2 ( \tens{\lambda} )$, but also proportional to $\tau \tens{\lambda}^2$, $\omega \tens{\lambda}^2$, ${\rm{\bf p}}^2 \tens{\lambda}^2$ and $( \tens{\lambda}^2 )^2$ are generated. Second, the basic task in computing proportional to $K_l ( \tens{\lambda} )$ is to determine the currents induced by the external current in the conducting propagators.

\section{Details on the two-loop diagrams}
\label{app:calculations2}
Here we sketch the computation of conducting two-loop diagrams with insertions. We restrict ourselves to a few examples. The techniques presented for these examples can then straightforwardly be adapted to the remaining diagrams. For briefness, we will exclusively consider those parts of the diagrams proportional to $K_l ( \tens{\lambda} )$. Moreover, we set external momenta and frequencies equal to zero.

At first we consider diagram $\mbox{H}$. We start by determining the currents flowing through the conducting propagators. Kirchhoff's law~(\ref{kirchhoff}) applies to the 4 vertices of the diagram. This allows us to eliminate 3 of the 5 unknown currents (one of the vertices is inactive with respect to this purpose since the external current $\tens{\lambda}$ must be conserved). The potential drop around closed loops is zero. Hence we can eliminate the two remaining unknown currents via the variation principle~(\ref{variationPrinciple2}) and express all currents flowing through conducting propagators in terms of the times and $\tens{\lambda}$. The momentum integrations are straightforward. They can be done by using the saddle point method which works exact here since the momentum dependence is purely quadratic. After carrying out the momentum integration we have
\begin{widetext}
\begin{eqnarray}
\label{H1}
&& \mbox{H}_{\mathcal{O}^{(l)}} = -w^l K_l \left( \vec{\lambda} \right) \frac{\rho g^4}{2} \frac{1}{(4\pi)^{d_\perp}} \int_0^\infty dt_1 dt_2 dt_3 \exp \left[ -\tau \left( 2t_1 +2t_2 +3t_3 \right) \right]  \frac{1}{(2t_3)^{d_\perp /2}}  
\nonumber \\
&& \times \,
\frac{1}{(2t_3)^{d_\perp /2}} \, \frac{1}{\left[ 2t_1 + 2t_2 + \frac{3}{2}t_3 \right]^{1+d_\perp /2}} \, \Bigg\{ \left( t_1 + t_2 \right) \left[ \frac{t_1 + t_2 + t_3}{2t_1 + 2t_2 + \frac{3}{2}t_3}  \right]^n 
\nonumber \\
&& + \,
\left( t_1 + t_2 + t_3 \right) \left[ \frac{t_1 + t_2 + \frac{1}{2} t_3}{2t_1 + 2t_2 + \frac{3}{2}t_3}  \right]^n +  2 t_3 \left[ \frac{1}{2} \frac{t_1 + t_2 + t_3}{2t_1 + 2t_2 + \frac{3}{2}t_3}  \right]^n \Bigg\} \, ,
\end{eqnarray}
\end{widetext}
where $n=2l$. Upon doing a little algebra we rewrite Eq.~(\ref{H1}) as 
\begin{eqnarray}
\label{H2}
&& \mbox{H}_{\mathcal{O}^{(l)}} = -w^l K_l \left( \vec{\lambda} \right) \frac{\rho g^4}{2} \Big\{ 2 I_1 + 2 I_2 + 2^{1-n} I_3 + I_4 \Big\} \, ,
\nonumber \\
\end{eqnarray}
where we have introduced the abbreviations
\begin{eqnarray}
&&I_1 = \frac{1}{(4\pi)^{d_\perp}} \int_0^\infty dt_1 dt_2 dt_3 \exp \left[ -\tau \left( 2t_1 +2t_2 +3t_3 \right) \right]
\nonumber \\
&&\times \,
 \frac{t_1 \left[ t_1 + t_2 + t_3 \right]^n}{(2t_3)^{d_\perp /2} \left[ 2t_1 + 2t_2 + \frac{3}{2}t_3 \right]^{n+d_\perp /2}} \, ,
\end{eqnarray}
\begin{eqnarray}
&&I_2 = \frac{1}{(4\pi)^{d_\perp}} \int_0^\infty dt_1 dt_2 dt_3 \exp \left[ -\tau \left( 2t_1 +2t_2 +3t_3 \right) \right] 
\nonumber \\
&&\times \,
 \frac{t_1 \left[ t_1 + t_2 + \frac{1}{2} t_3 \right]^n}{(2t_3)^{d_\perp /2} \left[ 2t_1 + 2t_2 + \frac{3}{2}t_3 \right]^{n+d_\perp /2}} \, ,
\end{eqnarray}
\begin{eqnarray}
&&I_3 = \frac{1}{(4\pi)^{d_\perp}} \int_0^\infty dt_1 dt_2 dt_3 \exp \left[ -\tau \left( 2t_1 +2t_2 +3t_3 \right) \right] 
\nonumber \\
&&\times \,
 \frac{t_3 \left[ t_1 + t_2 + t_3 \right]^n}{(2t_3)^{d_\perp /2} \left[ 2t_1 + 2t_2 + \frac{3}{2}t_3 \right]^{n+d_\perp /2}} \, ,
\end{eqnarray}
\begin{eqnarray}
&&I_4 = \frac{1}{(4\pi)^{d_\perp}} \int_0^\infty dt_1 dt_2 dt_3 \exp \left[ -\tau \left( 2t_1 +2t_2 +3t_3 \right) \right] 
\nonumber \\
&&\times \,
 \frac{t_3 \left[ t_1 + t_2 + \frac{1}{2} t_3 \right]^n}{(2t_3)^{d_\perp /2} \left[ 2t_1 + 2t_2 + \frac{3}{2}t_3 \right]^{n+d_\perp /2}} \, .
\end{eqnarray}
Now consider $I_1$. The integrations can be simplified by changing variables: $t_1 \to \frac{1}{2} ty$, $t_2 \to \frac{1}{2} t (1-x-y)$ and $t_3 \to \frac{2}{3} tx$. This gives after doing the integration over $t$ and $y$
\begin{eqnarray}
&&I_1 =  \frac{2^{-n}}{24} \left( \frac{3}{4} \right)^{d_\perp /2}\frac{\Gamma (4-d_\perp)}{(4\pi)^{d_\perp}} \, \tau^{d_\perp -4} \int_0^1 dx \, x^{-d_\perp /2} 
\nonumber \\
&&\times \,
\left(  1-x \right)^2 \left( 1 + x \right)^{d_\perp -4} \left( 1 + \frac{1}{3} x \right)^n \, . 
\end{eqnarray}
The remaining integral over $x$ may be simplified by separating its divergent and convergent contributions via Taylor expansion:
\begin{eqnarray}
&&I_1 =  \frac{2^{-n}}{24} \left( \frac{3}{4} \right)^{d_\perp /2}\frac{\Gamma (4-d_\perp)}{(4\pi)^{d_\perp}} \, \tau^{d_\perp -4} \int_0^1 dx \, \Bigg\{ x^{-d_\perp /2}
\nonumber \\
&& + \, x^{1 -d_\perp /2} \left( -2 + d_\perp -4 + \frac{n}{3} \right)  
\nonumber \\
&& + \, x^{-2} \left[ \left(  1-x \right)^2 \left( 1 + \frac{1}{3} x \right)^n -1 +2x - \frac{n}{3}x \right] \Bigg\} \, . 
\nonumber \\
\end{eqnarray}
Carrying out the integration and expansion for small $\varepsilon$ then gives the result
\begin{eqnarray}
&&I_1 =  \frac{3 \cdot 2^{-n}}{128} \, \frac{G_\varepsilon^2}{\varepsilon} \, \tau^{-\varepsilon} \Bigg\{ - \frac{4}{\varepsilon} + \frac{2\, n}{3\, \varepsilon} -3 + F_{2,3} (n) 
\nonumber \\
&& 
- \, 2 F_{1,3} (n) - \frac{3}{n+1} + \frac{3}{n+1} \left( \frac{4}{3} \right)^{n+1} 
\nonumber \\
&&
- 2 \, \ln \left( \frac{4}{3} \right) + \frac{n}{3} \ln \left( \frac{4}{3} \right) \Bigg\} \, . 
\end{eqnarray}
Here, we have used the shorthand
\begin{eqnarray}
F_{m,l} (n) = \sum_{k=m}^n \binom{n}{k} \frac{l^{-k}}{k-m+1} \, . 
\end{eqnarray}
$I_2$, $I_3$, and $I_4$ can be evaluated in the same fashion as $I_1$. Thus, we merely state the results:
\begin{eqnarray}
&&I_2 =  \frac{3 \cdot 2^{-n}}{128} \, \frac{G_\varepsilon^2}{\varepsilon} \, \tau^{-\varepsilon} \Bigg\{ - \frac{4}{\varepsilon} - \frac{2\, n}{3\, \varepsilon} -3 + F_{2,-3} (n) 
\nonumber \\
&&
- \, 2 F_{1,-3} (n) + \frac{3}{n+1}  - \frac{3}{n+1} \left( \frac{4}{3} \right)^{n+1} 
\nonumber \\
&&
- \, 2 \ln \left( \frac{4}{3} \right) - \frac{n}{3} \ln \left( \frac{4}{3} \right) \Bigg\} \, , 
\end{eqnarray}
\begin{eqnarray}
&&I_3 =  \frac{2^{-n}}{16} \, \frac{G_\varepsilon^2}{\varepsilon} \, \tau^{-\varepsilon} \Bigg\{ \frac{2}{\varepsilon} + F_{1,3} (n) 
\nonumber \\
&&
+ \, \frac{3}{n+1} - \frac{3}{n+1} \left( \frac{4}{3} \right)^{n+1} + \ln \left( \frac{4}{3} \right) \Bigg\} \, , 
\end{eqnarray}
\begin{eqnarray}
&&I_4 =  \frac{2^{-n}}{16} \, \frac{G_\varepsilon^2}{\varepsilon} \, \tau^{-\varepsilon} \Bigg\{ \frac{2}{\varepsilon} + F_{1,-3} (n) 
\nonumber \\
&&
- \, \frac{3}{n+1} + \frac{3}{n+1} \left( \frac{4}{3} \right)^{n+1} + \ln \left( \frac{4}{3} \right) \Bigg\} \, .
\end{eqnarray}
Upon collecting, we obtain for diagram $\mbox{H}$ the final result
\begin{widetext}
\begin{eqnarray}
\label{H3}
&& \mbox{H}_{\mathcal{O}^{(l)}} = -w^l K_l \left( \vec{\lambda} \right) \frac{\rho g^4}{2} \frac{3 \cdot 2^{-n}}{128} \, \frac{G_\varepsilon^2}{\varepsilon} \, \tau^{-\varepsilon}  \Bigg\{ - \frac{16}{3\, \varepsilon} - 6 +  F_{2,3} (n) + F_{2,-3} (n) 
\nonumber \\
&&
- \, 2 F_{1,3} (n) - \frac{2}{3} F_{1,-3} + \frac{3}{n+1} \left( \frac{4}{3} \right)^{n+1} + \frac{1}{n+1} \left( \frac{2}{3} \right)^{n+1} - \frac{4}{n+1} - \frac{8}{3} \ln \left( \frac{4}{3} \right) 
\nonumber \\
&&
+ \, 2^{1-n} \Bigg[ \frac{8}{3\, \varepsilon} + \frac{4}{3} F_{1,3} (n) + \frac{4}{n+1} - \frac{4}{n+1} \left( \frac{4}{3} \right)^{n+1} + \frac{4}{3} \ln \left( \frac{4}{3} \right) \Bigg] \Bigg\} \, .
\end{eqnarray}
\end{widetext}
As a further example we now treat diagram $\mbox{C}$. We employ once more our calculation scheme and determine for each conducting propagator the induced external current. This provides us with the noise cumulants of the diagram and leads to
\begin{eqnarray}
\label{C1}
&& \mbox{C}_{\mathcal{O}^{(l)}} = -w^l K_l \left( \vec{\lambda} \right) \rho g^4 \big\{ I_5 + 2 I_6 + 2 I_7 \big\} \, ,
\end{eqnarray}
where we have used the abbreviations
\begin{eqnarray}
&&I_5 = \frac{1}{(4\pi)^{d_\perp}} \int_0^\infty dt_1 dt_2 dt_3 \, \exp \left[ -\tau \left( 2t_1 +2t_2 +3t_3 \right) \right]
\nonumber \\
&& \times \,
\frac{t_3^{n+1}\left[ t_1 + t_2 + t_3 \right]^n}{\left[4 (t_1 + t_3)(t_2 + t_3) - t_3^2\right]^{n+d_\perp/2}}
\, ,
\end{eqnarray}
\begin{eqnarray}
I_6 &=& \frac{1}{(4\pi)^{d_\perp}} \int_0^\infty dt_1 dt_2 dt_3 \, \exp \left[ -\tau \left( 2t_1 +2t_2 +3t_3 \right) \right]
\nonumber \\
&& \times \,
\frac{t_1 \left[ 2 t_1 t_2 + 2 t_1 t_3 + 3 t_2 t_3 + 2 t_3^2\right]^2}{\left[4 (t_1 + t_3)(t_2 + t_3) - t_3^2 \right]^{n+d_\perp/2}}
\, ,
\end{eqnarray}
\begin{eqnarray}
I_7 &=& \frac{1}{(4\pi)^{d_\perp}} \int_0^\infty dt_1 dt_2 dt_3 \, \exp \left[ -\tau \left( 2t_1 +2t_2 +3t_3 \right) \right]
\nonumber \\
&& \times \,
\frac{\left( t_1 + t_3 \right)\left[ 2 t_1 t_2 + 2 t_1 t_3 + t_2 t_3 +  t_3^2 \right]^n}{\left[4 (t_1 + t_3)(t_2 + t_3) - t_3^2 \right]^{n+d_\perp/2}}
\, .
\end{eqnarray}
Exemplarily we drill into the integral $I_5$. It can be simplified by changing variables according to $t_1 \to t(x-1)$, $t_2 \to t(y-1)$, and $t_3 \to t$. Upon carrying out the integration over $t$ we obtain
\begin{eqnarray}
I_5 &=& \frac{\Gamma ( 4-d_\perp )}{(4\pi)^{d_\perp}} \tau^{d_\perp -4} \int_1^\infty dx dy 
\nonumber \\
&& \times \, \frac{\left[ 2x + 2y  -1 \right]^{d_\perp -4} \left[ x + y -1 \right]^n}{\left[ 4 xy -1 \right]^{n + d_\perp /2}} \, .
\end{eqnarray}
A further simplification can be achieved by rearranging the remaining integrations as\begin{eqnarray}
\label{blind}
I_5 &=& \frac{\Gamma ( 4-d_\perp )}{(4\pi)^{d_\perp}} \tau^{d_\perp -4} 2^{-7} \int_1^\infty dx \int_1^x dy
 \nonumber \\
&& \times \, \frac{\left[ x + y  - \frac{1}{2} \right]^{d_\perp -4} \left[ x + y -1 \right]^n}{\left[ xy - \frac{1}{4} \right]^{n + d_\perp /2}}
\, .
\end{eqnarray}
We learn from Eq.~(\ref{blind}) that $I_5$ is convergent for $d_\perp \leq 4$. Hence it is legitimate to evaluate it directly at $d_\perp = 4$. In contrast to the integrals constituting diagram H we were not able to evaluate $I_5$ for arbitrary $n$. The technical difficulty is the binomial appearing in the denominator of the integrand of $I_5$. For $n$ not too large, however, the number of terms of this binomial ($2^{n+2}$ at $d_\perp = 4$) is manageable and one can at least carry out the integrations for each reasonable $n$ separately. We refrain from stating all the results because this would be rather space consuming. We annotate that $I_6$ and $I_7$ can be treated in a similar fashion as $I_5$, except that these calculations are somewhat more tedious. $I_6$ and $I_7$ are not convergent like $I_5$ so that in practice one has to separate divergent and convergent contributions as it was demonstrated in considering diagram H.
%\end{widetext}

%references


\begin{thebibliography}{}
\bibitem{footnote1}
Sending $E$ and $D$ together to zero does no harm but it brings no benefit either.
\end{thebibliography}
\end{document}